\RequirePackage{fix-cm}
\documentclass[twocolumn]{svjour3}          
\smartqed  
\usepackage{graphicx}
\usepackage{balance}
\usepackage{algorithm}
\usepackage{algpseudocode}
\usepackage{amsfonts}
\usepackage{soul}
\usepackage{marginnote}
\usepackage{enumitem}
\usepackage{verbatim}
\usepackage{pifont}
\usepackage{float}\usepackage{stfloats}
\usepackage{color,soul}
\usepackage[T1]{fontenc}
\usepackage{url}
\usepackage{paralist}
\usepackage{caption}
\usepackage{subcaption}
\usepackage{theorem}
\theoremstyle{definition}
\newtheorem{exmp}{Example}
\newcommand{\Paragraph}[1]{\vspace{-0.02in}\smallskip\noindent{\bf #1.}}
\newcommand{\blue}[1]{\textcolor{blue}{#1}}

\begin{document}

\title{Coconut: Sortable Summarizations for Scalable Indexes over Static and Streaming Data Series
}

\author{
	%
	%
	\alignauthor
\\
	\affaddr{FORTH-ICS}\\
	\email{\large kondylak@ics.forth.gr}
	\alignauthor
	Niv Dayan\\
	\affaddr{Harvard University}\\
	\email{\large dayan@seas.harvard.edu}
	\and
	\alignauthor Kostas Zoumpatianos\\
	\affaddr{Harvard University}\\
	\email{\large kostas@seas.harvard.edu}
	\alignauthor Themis Palpanas\\
	\affaddr{Paris Descartes University}\\
	\email{\large themis@mi.parisdescartes.fr}
}

\author{Haridimos Kondylakis         \and
        Niv Dayan  \and
        Kostas Zoumpatianos  \and
        Themis Palpanas
}

\institute{H. Kondylakis \at
	FORTH-ICS \\
	\email{kondylak@ics.forth.gr}           
	\and
	N. Dayan \at
	Harvard University \\
	\email{dayan@seas.harvard.edu}           
	\and
	Kostas Zoumpatianos \at
	Harvard University \\
	\email{kostas@seas.harvard.edu}           
	\and
	Themis Palpanas \at
	Paris Descartes University \\
	\email{themis@mi.parisdescartes.fr}           
}


\date{Received: date / Accepted: date}

\maketitle

\sloppy
\begin{abstract}

Many modern applications produce massive streams of data series that need to be analyzed, requiring efficient similarity search operations. 
However, the state-of-the-art data series indexes that are used for this purpose do not scale well for massive datasets in terms of performance, or storage costs.
We pinpoint the problem to the fact that existing summarizations of data series used for indexing cannot be sorted while keeping similar data series close to each other in the sorted order.
To address this problem, we present Coconut, the first data series index based on sortable summarizations, and the first efficient solution for indexing and querying streaming series.
The first innovation in Coconut is an inverted, sortable data series summarization that organizes data series based on a z-order curve, keeping similar series close to each other in the sorted order.
As a result, Coconut is able to use bulk-loading and updating techniques that rely on sorting to quickly build and maintain a contiguous index using large sequential disk I/Os.
We then explore prefix-based and median-based splitting policies for bottom-up bulk-loading, showing that median-based splitting outperforms the state of the art, ensuring that all nodes are densely populated. Finally, we explore the impact of sortable summarizations on variable size window queries, showing that they can be supported in the presence of updates through efficient merging of temporal partitions. 
Overall, we show analytically and empirically that Coconut dominates the state-of-the-art data series indexes in terms of construction speed, query speed, and storage costs. 


\end{abstract}

\section{Introduction}
Many scientific and business applications today produce  massive collections and streams of data series\footnote{Informally, a \emph{data series}, or \emph{data sequence}, is an ordered sequence of data points. If the dimension that imposes the ordering of the sequence is time then we talk about \emph{time series}, though a series can also be defined over other measures (e.g., angle in radial profiles in astronomy, mass in mass spectroscopy, position in genome sequences, etc.). For the rest of this paper, we are going to use the terms \emph{data series} and \emph{sequence} interchangeably.} and need to analyze them, requiring the efficient execution of  similarity search, or nearest neighbor operations, over either the entire dataset, or variable-sized windows of the incoming data. 
Example applications range across the domains of audio~\cite{KashinoSM99}, images~\cite{YeK09}, finance~\cite{Shasha99}, telecommunications~\cite{humanbehaviorpatterns,DBLP:conf/edbt/MirylenkaCPPM16}, environmental monitoring~\cite{zzz}, scientific data~\cite{HuijseEPPZ14,url:adhd,VALMOD}, and others.

As the price of digital storage continues to plummet, the volume of data series collections grows,  driving the need for the development of efficient sequence management systems~\cite{DBLP:journals/sigmod/Palpanas15,DBLP:conf/ieeehpcs/Palpanas17,KostasThemisTalkICDE}. 
For the specific problem of sequence similarity search, searching for a nearest neighbor by traversing the entire dataset for every query quickly becomes intractable as the dataset size increases.
Consequently, multiple data series indexing techniques have been proposed over the past decade to organize data series based on similarity~\cite{DBLP:conf/sofsem/Palpanas16,lernaeanhydra}. 
The state-of-the-art approach is to index data series based on smaller summarizations that approximate the distances among data series.
This enables pruning large parts of the dataset that are guaranteed to not contain the nearest neighbor, and thereby these indexes significantly improve query speed.

Large data series collections and indexes that span hundreds of gigabytes to terabytes~\cite{url:adhd,url:sds,DBLP:journals/pvldb/PelkonenFCHMTV15} must reside in slow secondary storage devices for cost-effectiveness.
This poses a set of challenges for data series indexes. (1) They must support construction, updates  and queries using I/O efficient access patterns. (2) They must take up as little storage space as possible to be cost-effective and to minimize the physical space that queries traverse. (3) They must utilize the limited I/O bandwidth effectively by narrowing a query's search not only spatially but also temporally to the window size that is most appropriate for a given application.  



\Paragraph{Unsortable Summarizations}
In this paper, we show that the state-of-the-art data series indexes are designed in a manner that prevents them from  meeting the above challenges.
We pinpoint the problem to the fact that the summarizations, used as keys by data series indexes, are unsortable.
Existing summarizations~\cite{Lin2003b,Chakrabarti2002} partition and tokenize data series into multiple (independent) segments that are laid out in the summarized representation based on their original order within the data series; thus, sorting based on these summarizations would place together data series that are similar in terms of their beginning, i.e., the first segment, yet arbitrarily far in terms of the rest of the segments\footnote{This is analogous to sorting points in a multi-dimensional space based on one dimension.}. Hence, existing summarizations cannot be sorted while keeping similar data series next to each other in the sorted order. This leads to the following two problems. 



\Paragraph{Problem 1: Top-Down Insertions} The first problem is that traditional algorithms for efficiently bulk-loading and updating a database index cannot be used because they rely on being able to sort the data. 
Instead, state-of-the-art data series indexes perform bulk-loading and updates using top-down in-place insertions and splitting nodes as they fill up~\cite{DBLP:conf/sofsem/Palpanas16,isax2plus,ZoumpatianosIP16}. This approach leads to many small random I/Os to secondary storage that slow down both construction speed and updating during runtime. Moreover, the resulting nodes (after many splits) are non-contiguous in storage, meaning that querying also involves many slow random I/Os. 

Relying on top-down insertions also prevents data-series indexes from being able to temporally partition the data to enable efficient queries over variable-sized windows. 
The reason is that batched updates are periodically applied to the complete data structure through in-place split operations.
While this choice facilitates queries that touch the entire history of the data, the absence of temporal partitioning penalizes queries that need to touch smaller parts of the history.
Moreover, no matter the window size, pending updates are always applied in an inefficient manner, as existing indexes do not support merge-sort operations.
While various solutions~\cite{Camerra2010,isax2plus} have been proposed to partition pending updates to touch independent subsets of the index, still all temporal partitions are merged using top-down insertions, which are prohibitively expensive.

\Paragraph{Problem 2: Prefix-Based Node-Splitting} The second problem is that it is not possible to sort and thereby split data series evenly across nodes (i.e., using the median value as a splitting point). Instead, state-of-the-art data series indexes divide data series across nodes based on common prefixes across all segments. As a result, it is impossible for entries that do not share a common prefix in one or more of the segments to reside in the same node. We show that this leads to most nodes being nearly empty (i.e., their  fill-factor is low, which translates to an increased number of leaves). This slows down query speed and amplifies storage costs.

\Paragraph{Our Solution: Sortable Summarizations and Coconut}
To address these problems, we show how to transform existing data series summarizations into \textit{sortable summarizations}.
The core idea is interweaving the bits that represent the different segments, such that the more significant bits across all segments precede all less significant bits.
As a result, we describe the first technique for sorting data series based on their summarizations: the series are positioned on a z-order curve~\cite{morton1966}, in a way that similar data series are close to each other.

Moreover, we show that indexing based on sortable summarizations has the same ability as existing summarizations to prune parts of the index that do not contain the nearest neighbor, while it offers three additional benefits: it enables (i) efficiently bulk-loading and updating the index, (ii) packing data series more densely into nodes, and (iii) efficient merging of temporal partitions to allow variable-sized window queries. Furthermore, we show that using sortable summarizations enables data series indexes to leverage a wide range of indexing infrastructure.


We further introduce the \textbf{Co}mpact and \textbf{Con}tiguous Seq\textbf{u}ence Infras\textbf{t}ructure (Coconut).
Coconut is a novel data series indexing infrastructure that organizes data series based on \emph{sortable} summarizations.
It supports bulk-loading techniques and log-structured updates to enable maintaining a contiguous index. This eliminates random I/O during construction, updating \emph{and} querying. 
Furthermore, Coconut is able to split data series across nodes by sorting them and using the median value as a splitting point, leading to data series being packed more densely into leaf nodes (i.e., at least half full). 

In order to study the design space and isolate the impact of the different design decisions, we first introduce two variants: Coconut-Trie and Coconut-Tree, which split data series across nodes based on common prefixes and median values, respectively.
We show that Coconut-Trie dominates the state-of-the-art in terms of query speed because it creates contiguous leaves. We further show that Coconut-Tree dominates Coconut-Trie and the state-of-the-art in terms of construction speed, query speed \emph{and} storage overheads because it creates a contiguous, balanced index that is also densely populated. We then introduce Coconut-LSM to support efficient log-structured updates and variable-size window queries over different windows of the data based on recency. Overall, we show across a wide range of workloads and datasets that Coconut-Tree improves both construction speed and storage overheads by one order of magnitude and query speed by two orders of magnitude relative to the state-of-the-art. We further show that Coconut-LSM supports updates without degrading query throughput, and that it is able to narrow the search scope temporally. This improves query throughput by a further 2-3 orders of magnitudes in our experiments for queries over recent data.

Our contributions are summarized as follows.
\begin{compactitem} 
	\item We show that existing data series summarizations cannot be sorted in a straightforward way. Consequently, state-of-the-art data series indexes cannot efficiently bulk-load and pack data densely into nodes, leading to large storage overheads and performance bottlenecks for both index construction and query answering, when dealing with very large data series collections.
	\item We introduce a \emph{sortable} data series summarization that keeps similar data series close to each other in the sorted order, and preserves the same pruning power as existing summarizations. We show how sortability enables new design choices for data series indexes,  thereby opening up infrastructure possibilities that were not possible in the past.
    \item We introduce Coconut-Trie that exploits sortable summarizations for prefix-based bulk-loading of existing state-of-the-art indexes, leading to improvements at querying time performance. 
	\item We present Coconut-Tree, which employs median-based bulk-loading to quickly build the index and to restrict space-amplification, by enabling entries that do not share a common prefix to be in the same node.	
	\item We introduce Coconut-LSM to enable efficient similarity search over variable-sized windows in the presence of updates. 
	\item Our experimental evaluation with a variety of synthetic and real datasets demonstrates that Coconut-Tree and Coconut-LSM strictly dominate existing state-of-the-art indexes in terms of both construction speed and storage overheads by one order of magnitude, and query speed by two orders of magnitude. We further show that Coconut-LSM dominates the state-of-the-art by orders of magnitude in the presence of insertions for queries over recent data. 
\end{compactitem} 

A preliminary version of this paper has appeared in VLDB~\cite{DBLP:journals/pvldb/KondylakisDZP18}. This version extends the previous one by introducing Coconut-LSM for efficient similarity search in the presence of updates, and presents the first efficient solution for indexing and querying streaming sets, along with the corresponding experiments.
We have also developed a system that implements the ideas described in this paper~\cite{DBLP:conf/sigmod/KondylakisDZP19}.

\section{Preliminaries and Related Work} 
\label{sec:background}

\Paragraph{Data Series}
Measuring data that fluctuate over a dimension is a very frequent scenario in a large variety of domains and applications.
Such data are commonly called data series or sequences. The dimension over which they fluctuate can range from time, angle or position to any other dimension. They can be measured at either fixed or variable intervals.
\begin{definition}
Formally, a data series $s=\{r_1, ..., r_n\}$ is defined as an ordered set of recordings, where each $r_i = <p_i, v_i>$ describes a value $v_i$ corresponding to a position $p_i$.
\end{definition}


\Paragraph{Nearest Neighbor Search}
Analysts perform a wide range of data mining tasks on data series including clustering~\cite{keogh1998,liao2005,rodrigues2008,rakthanmanon2011}, classification and deviation detection~\cite{Shieh2009,Shandola2009}, frequent pattern mining~\cite{DBLP:journals/datamine/MueenKZCWS11,DBLP:journals/tkdd/GrabockaSS16}, and more.
Existing algorithms for executing these tasks rely on performing fast similarity search across the different data series.
Thus, efficiently processing nearest neighbor (NN) queries is crucial for speeding up the aforementioned tasks.
NN queries are formally defined as follows.

\begin{definition}
Given a set of data series $\bold{S} \subseteq \mathcal{S}$, where $\mathcal{S}$ is the set of all possible data series, a query data series $s_q \in \mathcal{S}$ and a distance function $d(\bullet, \bullet) : \mathcal{S} \times \mathcal{S} \rightarrow \mathbb{R}$, a nearest neighbor query is defined as:
$$
nn_{d(\bullet, \bullet)}(s_q, \bold{S}) = s_i \in \bold{S} : d(s_i, s_q) \leq d(s_j, s_q) \forall s_j \neq s_i \in \bold{S}.
$$
\end{definition}

Common distance metrics for comparing data series include Euclidean Distance (ED) and dynamic time warping (DTW).
While DTW is better for most data mining tasks, the error rate using ED converges to that of DTW as the dataset size grows~\cite{Ratanamahatana2005,Xiaopeng2006,Shieh2008}.
Therefore, data series indexes for massive datasets use ED as a distance metric~\cite{Shieh2009,Shieh2008,Zoumpatianos2014,Zoumpatianos2015rinse,ZoumpatianosIP16}, though simple modifications can be applied to make them compatible with DTW~\cite{Shieh2008,Kate2016}. Euclidean distance is computed as the sum of distances between pairs of aligned points in sequences of the same length, where normalizing the sequences for alignment and length is a pre-processing step~\cite{Shieh2009,Shieh2008,Zoumpatianos2014,Zoumpatianos2015rinse,ZoumpatianosIP16}.
In all cases, data are z-normalized by subtracting the mean and dividing by the standard deviation (note that minimizing ED on z-normalized data is equivalent to maximizing their Pearson's correlation coefficient~\cite{MueenNL10}).



{\color{red}
}

\begin{figure}[tb]
	\center
	\includegraphics[width=0.9\columnwidth]{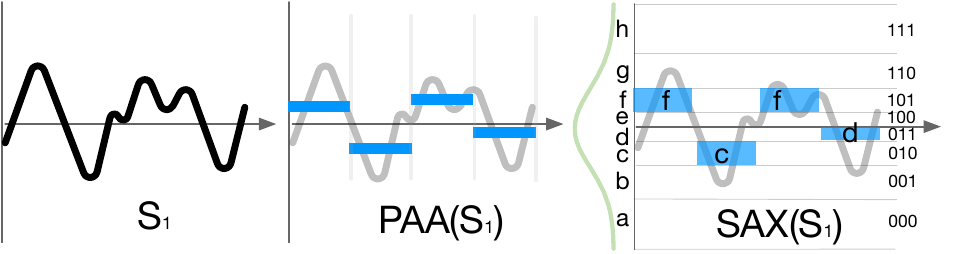}
	\caption{Example PAA and SAX summarizations.}\label{fig:sax}
\end{figure}


\Paragraph{Brute-Force Search}
The brute-force approach for evaluating nearest neighbor queries is by performing a sequential pass over the complete dataset\footnote{Note that recent state-of-the-art serial scan algorithms~\cite{rakthanmanon2012searching,DBLP:conf/icdm/MueenHE14} are only efficient for scenarios that involve nearest neighbor operations of a short query subsequence against a very long data series. On the contrary, in this work, we are interested in finding similarities in very large collections of short sequences.}.
However, as data series collections grow to terabytes~\cite{url:adhd,url:sds,DBLP:journals/pvldb/PelkonenFCHMTV15}, scanning the complete dataset becomes  performance bottleneck taking hours or more to complete. This is especially problematic in exploratory search scenarios, where batch execution of queries is impossible because the next query depends on the results of previous queries. 

\Paragraph{Data Series Summarizations} To mitigate this problem, various dimensionality reduction techniques have been proposed to transform data series into summarizations that enable approximating and lower bounding the distance between any two data series.  Examples include generic Discrete Fourier Transforms (DFT)~\cite{Agrawal1993,DBLP:conf/sigmod/FaloutsosRM94,Rafiei1997,Rafiei1998}, Piecewise Linear Approximation (PLA)~\cite{Keogh1997}, Singular Value Decomposition (SVD)~\cite{Korn97,RaviKanth1998}, Discrete Haar Wavelet Transforms (DHWT)~\cite{ChanF99,DBLP:conf/kdd/KashyapK11}, Piecewise Constant Approximation (PCA), and Adaptive Piecewise Constant Approximation (APCA)~\cite{Chakrabarti2002}, as well as data series specific techniques such as Piecewise Aggregate Approximation (PAA)~\cite{Keogh2000}, Symbolic Aggregate approXimation (SAX)~\cite{Lin2003} and the indexable Symbolic Aggregate approXimation ($i$SAX)~\cite{Shieh2008,Camerra2010}.
These smaller summarizations can be scanned and filtered~\cite{DBLP:conf/kdd/KashyapK11,Li1996}, or indexed and pruned~\cite{ZoumpatianosIP16,Shieh2009,Shieh2008,Zoumpatianos2014,Zoumpatianos2015rinse,Guttman1984,Assent2008,Wang2013,DBLP:conf/icdm/YagoubiAMP17,ulisse,conf/bigdata/peng18,ulissevldb,dpisaxjournal} to avoid accessing parts of the data that do not contain the nearest neighbor. 

\Paragraph{Clustering Approaches} Various clustering algorithms have been proposed for data series~\cite{kaufman2009,liao2005}, and such approaches can be used to facilitate nearest neighbor search. The general approach involves adapting distance measure between data series and using a clustering algorithm on top (e.g., K-means~\cite{macqueen1967}, K-shape~\cite{PaparrizosG15}, agglomerative \cite{kaufman2009}, etc.). Such algorithms require multiple passes over the data to build (e.g., to measure distances between all pairs of points as in agglomerative clustering, or to iteratively refine clusters with K-means). 
As a result, construction can take a very long time. 
In contrast, we focus on approaches based on indexable summarizations that are designed to lead to fast index construction, and thereby shorten the indexing-to-query time.





\Paragraph{Data Series Indexing with SAX}
We now discuss the state-of-the-art in data series indexing.
We concentrate on SAX summarizations \cite{Shieh2008,Lin2003}, which have been shown to outperform other summarizations in terms of pruning power using the same amount of bytes~\cite{ZoumpatianosLPG15}.
We illustrate the construction of a SAX summarization in Figure \ref{fig:sax}.

SAX first partitions the data series in equal-sized segments, and for each segment it computes its average value.
This is essentially a PAA summarization, and can be seen in Figure~\ref{fig:sax}(middle).
In a second step, it discretizes the value space by partitioning it in regions, whose size follows the normal distribution.
As a result, we have more regions corresponding to values close to 0, and less regions for the more extreme values (this leads to an approximately equal distribution of the raw data series values across the regions, since extreme values are less frequent than values close to 0 for z-normalized series).
A bit-code (or a symbol) is then assigned to every region.
The data series is then summarized by the sequence of symbols of the regions in which each PAA value falls.

In the example in Figure~\ref{fig:sax}, the data series $S_1$ becomes ``fcfd''.
This lossy representation requires much less space (typically in the order of 1\%) and reduces the number of dimensions from the number of points in the original series to the number of segments in the summarization (four in Figure~\ref{fig:sax}).

Data series indexes based on SAX rely on a multi-resolution indexable SAX representation (iSAX)~\cite{Shieh2008,Shieh2009} whereby every node corresponds to a common SAX prefix from across all segments. When a node fills up, the segment whose next unprefixed digit divides the resident data series most is selected for splitting the data series across two new nodes. $i$SAX 2.0~\cite{Camerra2010} and $i$SAX 2+~\cite{isax2plus} are variants that improve construction speed by storing all internal nodes in main memory and buffering access to leaf nodes.
ADS~\cite{Zoumpatianos2014,Zoumpatianos2015rinse,ZoumpatianosIP16} represents the state-of-the-art method and builds on these ideas by constructing an index based on the summarizations; the method then incorporates the raw data series into the index adaptively during query processing.

These indexes all share the following four performance problems. (1) If main memory is small relative to the raw data size, they incur many random I/Os due to swapping and early flushing of buffers. This significantly elongates construction time and updates for massive datasets. 
(2) The resulting leaf nodes after many splits are non-contiguous in secondary storage and therefore require many slow random I/Os to query. 
(3) Temporal partitioning to enable window queries over recent data cannot be performed efficiently, because different temporal partitions cannot be easily merged. This operation requires top-down entry-by-entry insertions, which lead to many small random I/Os.
(4) Data series that do not share common prefixes cannot reside in the same node, and so the leaf nodes in these indexes are in practice sparsely populated. This leads to significant storage overheads and slows down queries as they must traverse a greater physical area to access the same data.





	Our work follows the same high-level idea of indexing the data series based on a smaller summarization to enable pruning, though our work is the first to use sortable summarizations to speed up index construction, updating and querying and to restrict storage space. In all previous work, the index is constructed and maintained through top-down insertions that lead to many slow random I/Os and to a sparsely populated, non-contiguous and unbalanced index. Our work is the first to use fast bottom-up bulk-loading, log-structured updates, and median-based splitting to efficiently build and maintain a contiguous, balanced, and densely populated index. Note that our infrastructure can be used in conjunction with any summarization that represents a sequence as a multi-dimensional point, and so it is compatible with all main-stream summarizations~\cite{Shieh2008,Agrawal1993,DBLP:conf/sigmod/FaloutsosRM94,Rafiei1997,Rafiei1998,Keogh1997,Korn97,RaviKanth1998,ChanF99,DBLP:conf/kdd/KashyapK11,Chakrabarti2002,Camerra2010}.



\section{Problem: Unsortable Summarizations}
\label{sec:problem}

\begin{figure}[tb]
    \centering
    \includegraphics[width=1\columnwidth]{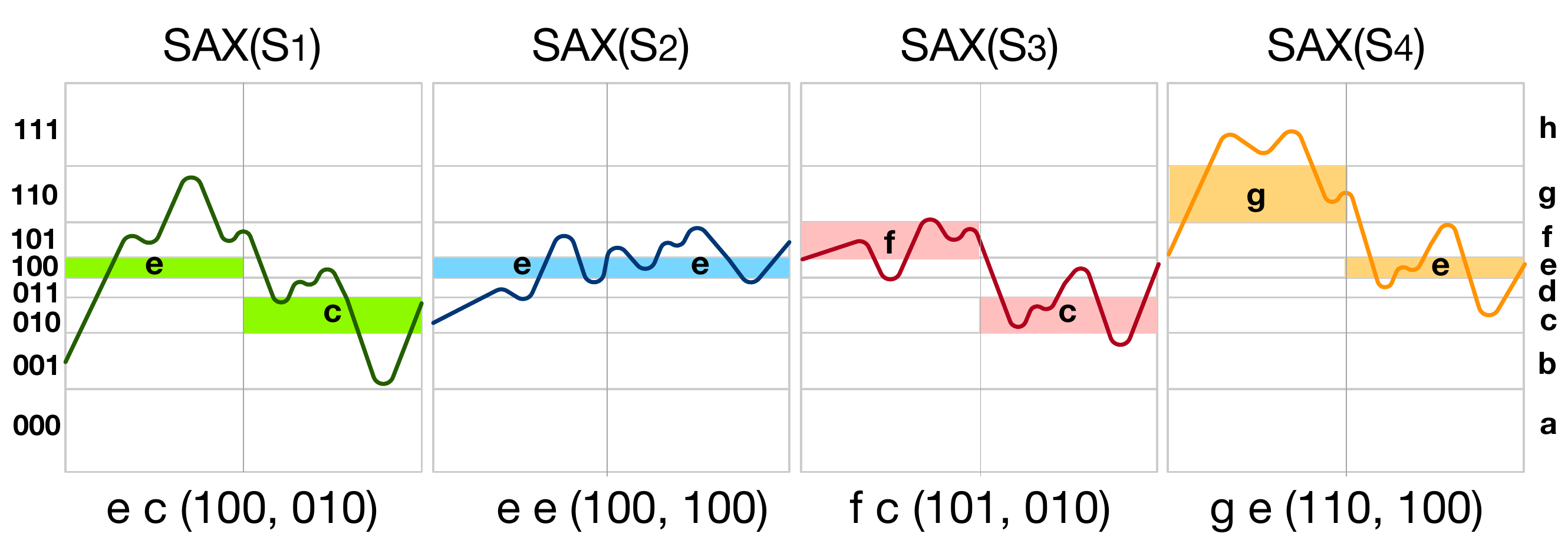}
    \caption{Sorting $i$SAX summarizations.}
    \label{fig:sorting}
\end{figure}

In this section, we describe why existing data series summarizations are not sortable, and we discuss the implications on index design, performance, and storage overheads. 

\Paragraph{Sorting summarizations}
Figure~\ref{fig:sorting} gives an example of sorting data series based on SAX summarizations.
There are four different data series with corresponding 2-character SAX words\footnote{Note that SAX words are typically longer to enable more precision; we use 2-character SAX words in this example for ease of exposition.}: $S_1 = ec$, $S_2 = ee$, $S_3 = fc$, and $S_4 = ge$.
Observe that $S_1$ is most similar to $S_3$, while $S_2$ is most similar to $S_4$ (apart from small differences in the first segments).
Sorting these summarizations lexicographically gives the order $S_1, S_2, S_3, S_4$: the data series that are most similar to each other are \emph{not} placed next to each other in the sorted order.
The reason is that existing summarizations lay out the segment representations sequentially, one by one.
Sorting based on such a representation would place next to each other data series that are similar in terms of their first segment, yet arbitrarily dissimilar in terms of the rest of the segments.
As a result, an index that is built by sorting data series based on existing summarizations would degenerate to scanning the full dataset for each query and would defeat the point of having an index.

It is important to note that even though we use SAX, the same observations hold for all other main-stream summarizations (discussed in Section~\ref{sec:background}).
This is because they all represent data series as multi-dimensional points.
As a result, they still suffer from the problem of poor lexicographical ordering, where sorting is based on arbitrarily ordering dimensions.
SAX was chosen in our work, since it has been shown to outperform other approaches in terms of quality~\cite{ZoumpatianosLPG15} and index performance~\cite{isax2plus,Zoumpatianos2014,Camerra2010}.

We next discuss how existing data series indexes overcome the inability to sort summarizations, and we analyze the impact on performance and storage overheads.




\begin{figure}[tb]
	\centering
  \includegraphics[width=0.9\columnwidth]{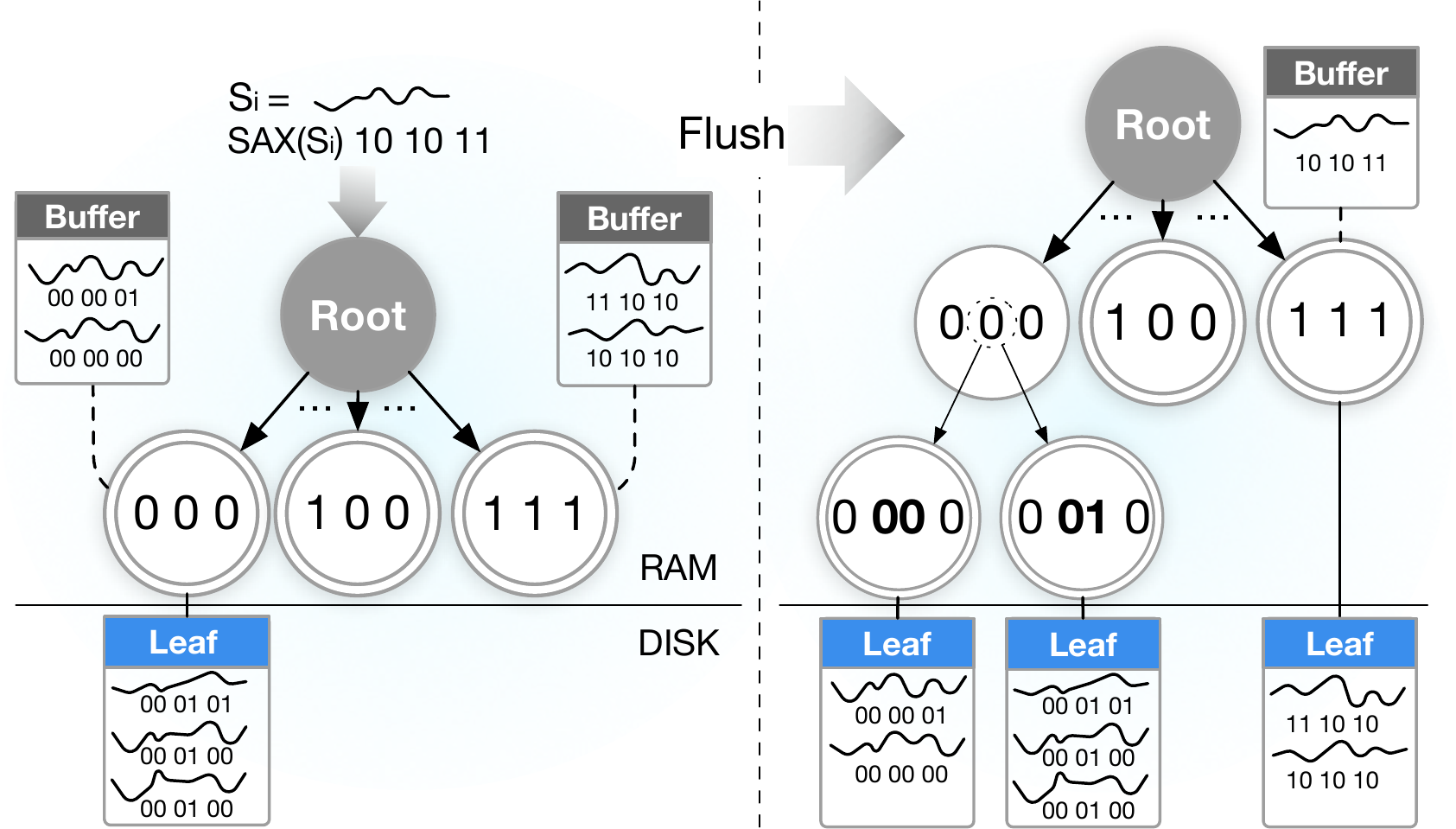}
  \caption{Indexing using $i$SAX 2.0.}
  \label{fig:isaxindexing}
\end{figure}
\begin{table}[ht]
\begin{center}
    \begin{tabular}{ | c | l | }
        \hline
        \textbf{Term} & \textbf{Definition} \\
        \hline
        $N$ & Total number of data series \\  \hline
        $B$ & Number of data series that fit into one disk block \\  \hline
        $M$ & Number of data series that fit into main memory  \\ 
        \hline
    \end{tabular}
    \caption{Table of terms}
    \label{tab:terms}
\end{center}
\end{table}

\subsection{Top-Down Insertions}

The standard approach for bulk-loading a database index (e.g., a B-Tree) relies on external sorting.
This approach cannot be used with existing data series summarizations, because they are not sortable.
Instead, state-of-the-art data series indexes perform top-down insertions~\cite{isax2plus,ZoumpatianosIP16,Wang2013}.
Here we analyze and compare their implications on performance and storage overheads. We analyze them in the disk access model \cite{Aggarwal1988}, which measures the runtime of an algorithm in terms of disk blocks transferred between main memory and secondary storage. The terms we use are in Table \ref{tab:terms}.

\Paragraph{The Current Approach: Top-Down Insertions}
Data series indexes are built and maintained using top-down insertions: each data series is inserted through the root node and trickles down to the appropriate leaf node~\cite{Shieh2009,Shieh2008}. Since the internal nodes are maintained in memory \cite{isax2plus,Camerra2010}, every top-down insertion involves at most three I/Os: one to read the appropriate leaf node, one to update it, and one to create a new leaf node in case the first one splits. The cost per insertion is therefore at most {\small $O(1)$} I/O, and so the cost of index construction is at most {\small $O(N)$} I/Os. As new leaf nodes are allocated wherever there is space on disk, adjacent nodes in the logical space are not necessarily continuous in storage.

State-of-the-art data series indexes strive to reduce construction cost by buffering insertions in main memory before flushing them to storage. This process is illustrated on Figure~\ref{fig:isaxindexing} for the $i$SAX 2.0 index.
The new series to be inserted, $S_i$, is translated to the $i$SAX word (10 10 11).
At the first level of the tree, data is split based on the first bit at each of the segments. 
As a result $S_i$ is buffered as a part of the (1 1 1) sub-tree.
In our example, all the buffers are full and so the new insertion causes them to flush and get consolidated with corresponding leafs in storage. 
During this operation, when a leaf node runs out of capacity, it creates two new children by increasing the number of bits used to represent one of the segments and divides the data series between them (we discuss this process in detail in Section~\ref{sec:splitting}).
The right side of Figure~\ref{fig:isaxindexing} shows an example where node (0 0 0) splits into two new nodes, (0 00 0) and (0 01 0).
The new leafs are allocated with free space to be able to absorb new insertions. 
With ample spatial locality in the insertion pattern, multiple entries in the buffer map onto a small set of {\small $\frac{M}{B}$} leaf nodes. Since the buffer flushes {\small $\frac{N}{M}$} times during index construction, the best-case construction cost with buffering is {\small $\frac{M}{B} \cdot \frac{N}{M} \in O(\frac{N}{B})$} I/O.  With little spatial locality, however, each entry from the buffer maps onto a different leaf node, thereby leading to a cost of {\small $M \cdot \frac{N}{M} \in O(N)$} I/O, the same as without buffering. Hence, buffering cannot in general alleviate the high index construction cost of top-down insertions, and it also cannot ensure that adjacent logical nodes are contiguous in storage.

{\color{blue}
}

\Paragraph{The Elusive Alternative: Bottom-up Insertions} Building an index on a batch of $N$ application insertions through external sorting comprises two phases: partitioning and merging.
The partitioning phase involves scanning the raw file in chunks that fit in main memory, sorting each chunk in main memory, and flushing it to secondary storage as a sorted partition. This amounts to two passes over the data. The merging phase involves merge-sorting all the different partitions into one contiguous sorted order, using one input buffer for each partition and one output buffer for the resulting sorted order.
Once the data is ordered, we build the index bottom-up.
Thus, the merging phase amounts to two additional passes over the data, and so external sorting involves overall four passes over the data.
This amounts to {\small $O(N/B)$} I/Os with a cost per insertion of {\small $O(1/B)$} I/O (the reason being that each I/O handles $B$ entries)\footnote{In fact this condition only holds as long as $M > \sqrt{N}$ \cite{DBLP:books/daglib/0011128}.
Since main memory is approximately two orders of magnitude more expensive than secondary storage, this condition holds in practice for massive datasets.}.


\Paragraph{Implications for Index Construction} The analysis in the disk access model above shows that external sorting dominates top-down insertions in terms of worst-case index construction cost because we only need to do a few passes amounting to {\small $O(N/B)$} I/Os rather than {\small $O(N)$}  random I/Os. Since a disk block $B$ is typically large relative to data elements, this amounts to a 1-2 order of magnitude difference in construction speed.

\Paragraph{Implications for Dynamic Insertions}
In a dynamic setting with ongoing insertions during runtime, every top-down insertion that takes place requires reading a target block from storage and rewriting it at a cost of {\small $O(1)$} I/O per insertion. 
The inability to sort the data means that data structures with better performance properties for ingesting insertions during runtime cannot be leveraged. For example, many modern write-optimized data structures buffer insertions and later sort-merge them multiple times while amortizing the overheads of sorting through large sort-merge operations. For example, the log-structured merge-tree (LSM-tree) has an I/O cost per insertion of {\small $O(\frac{log(N)}{B})$} as it merges each entry a logarithmic number of times, but the sort-merge operations allow us to divide this cost by the block size $B$.  A (traditional) data-series index, however, cannot sort-merge the data and so it would have to rely on top-down insertions to merge the runs thereby blowing up the insertion cost to {\small $O(log(N))$} and making the scheme impractical. Thus, write-optimized data structures are currently inapplicable for data series indexing.





\Paragraph{Implications for Window Queries}
Performing window queries requires creating temporal partitions of the data so that a query can skip partitions with older data that are not needed by the application. Existing data series indexes do not perform temporal partitioning, and so the cost for a window query over the last $X$ insertions only requires searching the whole index at a cost of {\small $O(\frac{N}{B})$} I/O. 
On the other hand, the log-structured merge tree, for instance, creates a logarithmic number of partitions of exponentially increasing sizes, and so performing a query with a selectivity of $s$ over data within the most recent window of $X$ insertions requires performing {\small $O(\frac{X \cdot r}{B})$} I/O, where $r$ is the size ratio across the different runs of LSM-tree \cite{Alsubaiee15}. The problem, as we just saw, is that using LSM-tree is that it blows up the cost of insertions by a logarithmic factor since the sort-merge operations cannot be performed efficiently. For this reason, such data structures that naturally temporally partition the data and offer support for window queries cannot be used, and as a result, window queries cannot be supported efficiently.

\Paragraph{Implications for  General Query Processing} Performing bulk-loading and insertions through external sorting has two performance advantages for subsequent query processing. Firstly, the sorted order can be written contiguously in secondary storage, meaning that queries can traverse leaves using large sequential I/Os rather than small random I/Os. Secondly, it is possible to pack data series as compactly as possible in nodes rather than leaving free space for future insertions. Immediately after bulk-loading, this saves storage costs and speeds up queries by reducing the physical space that a query must traverse by a factor of 2.

\Paragraph{Summary}
Overall, external sorting dominates top-down insertions in terms of both construction and query speed. The problem is that existing data series indexes cannot use external sorting as they cannot sort the data based on existing data series summarizations.

\subsection{Splitting Nodes} \label{sec:splitting}

Database indexes such as B-trees split nodes when they run out of capacity using the median value as a splitting point, whereas data series indexes use prefix-based splitting. We now describe these methods in detail and analyze their implications on performance and storage overheads. We again use the disk access model \cite{Aggarwal1988} to quantify storage overheads in terms of disk blocks.


\Paragraph{Prefix-Based Splitting} In state-of-the-art data series indexes, every node is uniquely identified by one prefix for every segment of the SAX representation, and all elements in the node or its subtrees have matching prefixes for all segments. When a leaf node runs out of capacity, we scan the summarizations and identify the segment whose next unprefixed bit divides the elements most. We create two new children nodes and divide the elements among them based on the value of this bit. The problem is that data is not guaranteed to be unevenly distributed across the nodes. In the worst-case, every node split divides the entries such that one moves to one of the new nodes and the rest move to the other, meaning that the index is unbalanced, most nodes contain only 1 entry, and so storage consumption is $O(N)$ disk blocks.


\Paragraph{Median-Based Splitting} Splitting a node using the median value involves sorting the data elements to identify the median, moving all elements to the right of this mid-point into a new node, and adding a pointer from the parent to the new node to ensure the index remains balanced. This approach ensures that every node is at least half full. As a result, the amount of storage space needed is at most double the size of the actual data. This amounts to $O(N/B)$ blocks. 


\Paragraph{Comparison} Prefix-based splitting results in an unbalanced index amplifies worst-case storage overheads relative to median-based splitting by a factor of $B$. Since exact query answering time is proportional to the number of leaf nodes in the index, it amplifies it  by the same factor. Overall, median-based splitting dominates prefix-based splitting, but we cannot use it in the context of data series indexing because existing summarizations are not sortable.

\section{Coconut}
\label{sec:bottomupindexconstruction}

In this section, we present Coconut in detail. Coconut is a novel data series indexing infrastructure that organizes data series based on sortable summarizations. As a result, Coconut indexes are able to use bulk-loading techniques based on sorting to efficiently build a contiguous index. Furthermore, they are able to divide data series among nodes based median values to ensure that the index is balanced and that all nodes are densely populated. Finally, Coconut indexes are able to leverage different data structures during runtime to support different read/write cost trade-offs, and they can optimize particularly well for streaming applications that require different temporal views over the data.

In Section~\ref{sec:sortable}, we show how to make existing summarizations sortable using a simple algorithm that interleaves the bits in a summarization such that all more significant bits from across all segments precede all less significant bits.
In Sections~\ref{sec:trie} and ~\ref{sec:tree},  we introduce Coconut-Trie and Coconut-Tree, respectively.
These data structures allow us to isolate and study the impact of the properties of contiguity and compactness on query and storage overheads. In Section ~\ref{sec:lsm}, we introduce Coconut-LSM, the first data series index that supports efficient, log-structured insertions during runtime.

\subsection{Sortable Summarizations}
\label{sec:sortable}

	Each data series summarization can be viewed as a point in multi-dimensional space, where each segment in the summarization represents a dimension. 
	The question is how to place points that are similar across all dimensions as close to each other as possible in storage so as to minimize disk access during similarity search.
	
	A well-known technique is to use a space-filling curve, which linearizes multi-dimensional data on storage while preserving locality. We illustrate an example in Figure~\ref{fig:sorting} with a z-order curve ~\cite{morton1966}, which linearizes data by using recursive Z shapes which allow proximal points to remain close to each other in the linearized order. 
	
	The standard technique for projecting entries into a Z-order is to (1) interleave the bit representation of all segments of an entry, and then (2) sorting the entries based on the inverted bit representation \cite{bayer1998ub,RamsakMFZEB00}. The intuition is that each dimension is represented as a bit string, whereon more significant bits carry more information, while smaller bits increase precision. 
	Conceptually, sorting data  is an operation that involves recursively dividing data entries based on the most significant bit into a hierarchy of sets, and then laying out the elements in the hierarchy in a depth-first order. Sorting the inverted summarizations therefore places more importance on co-locating entries that are similar across their most significant bits, and a decreasing amount of importance on being closer in terms of each segment’s lesser significant bits. 
	An implementation of this technique for data series is shown in Algorithm~\ref{sortablesdrs}, transforming existing summarization schemes into sortable ones. To the best of our knowledge we are the first to apply this into data series summarizations.
	
\begin{figure}[tb]
	\centering
	\includegraphics[width=0.8\columnwidth, angle=0]{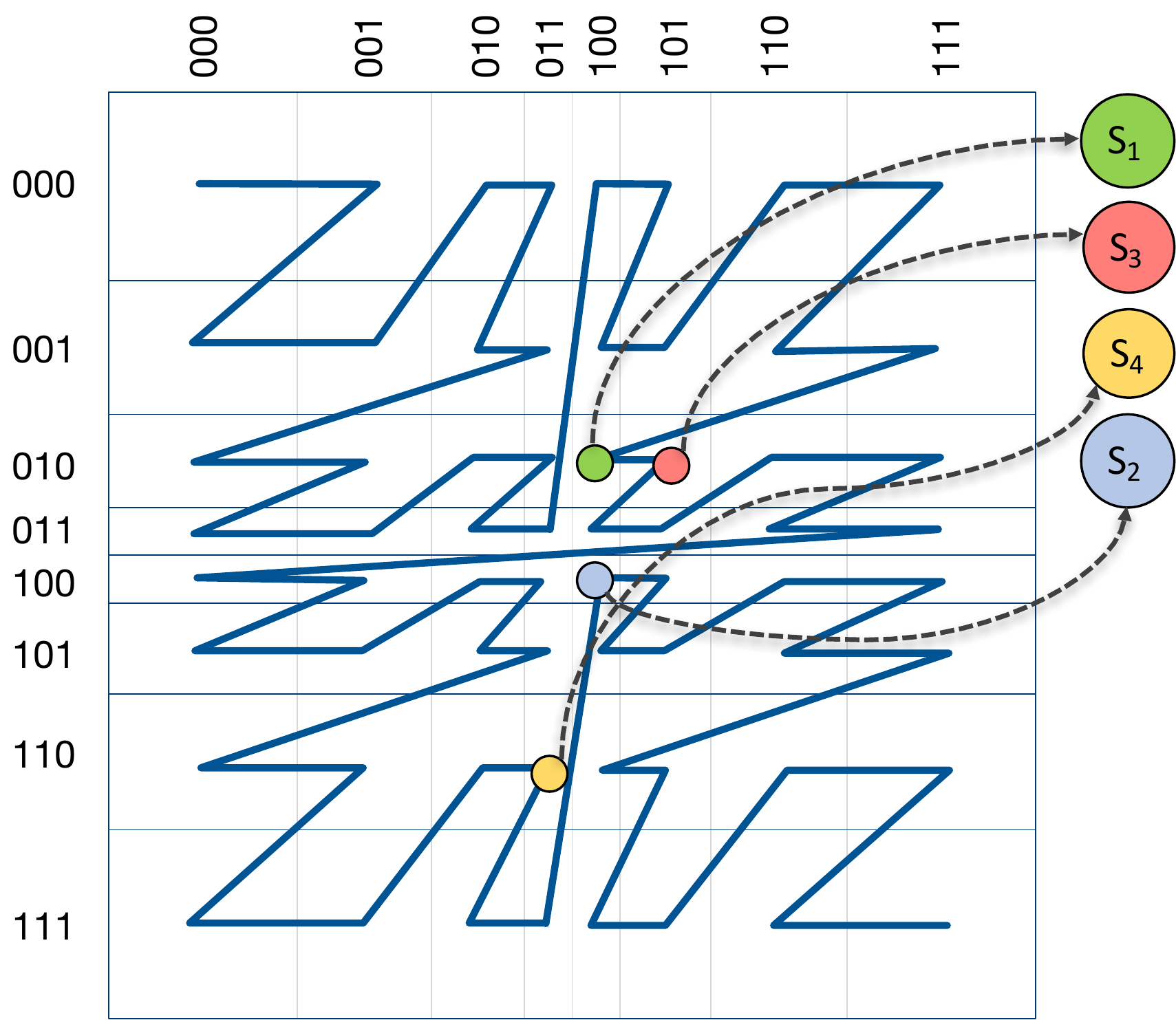}
	\caption{Z-ordered SAX summarization.}
	\label{fig:sortingz}
\end{figure}

	Figure~\ref{fig:sortingz} shows how to transform the four summarizations in Figure~\ref{fig:sorting} into sortable Z-ordered summarizations in two dimensions (for ease of illustration). 
	The technique applies to data with any number of segments/dimensions. 
	The figure also shows their linearized order along the z-ordered curve. As shown, the data series that are most similar to each other are indeed placed closest to each other (which is not the case when sorting them based on the original representation).

	Note that a sortable summarization contains the same amount of information as the original summarization, the only difference being that the bits are ordered differently.
	Hence, it is easy and efficient to switch back and forth between sortable summarizations and the original form, and we therefore do not lose anything in terms of the ability to prune the index during querying.

\Paragraph{New Infrastructure Opportunities} The ability to sort data series summarizations enables a plethora of new indexing infrastructure possibilities for data series indexes, ranging from read-optimized B-trees~\cite{Rao2000} to write-optimized LSM-trees~\cite{DBLP:journals/acta/ONeilCGO96} to adaptive structures that change performance characteristics based on workload~\cite{idreos2007,dayan2017}. Coconut-Trie, Coconut-Tree, and Coconut-LSM represent three points in this space that push upon the current state-of-the-art, though we expect that many more opportunities for specialization based on hardware and workload are possible.




\begin{algorithm}[tb]
\caption{Sortable Summarization}
\label{sortablesdrs}
\begin{algorithmic}[1]
	\small
\Procedure{invertSum}{Sum}
\For{each bit i of a segment in Sum}
    \For{each segment j}
         \State Add the i bit of segment j to SSum
    \EndFor
\EndFor
\State return SSum
\EndProcedure
\end{algorithmic}
\end{algorithm}


\subsection{Coconut-Trie}
\label{sec:trie}

We now present Coconut-Trie, a data series index that uses sortable summarizations to
construct a contiguous index using bulk-loading. Similarly to the state-of-the-art indexing schemes, Coconut-Trie divides data entries among nodes based on the greatest common prefix among all segments.
The advantage relative to the state-of-the-art is that
the resulting index is contiguous, meaning that queries do not issue random I/Os, but a large sequential I/O.

\Paragraph{Construction} The construction algorithm is shown in Algorithm~\ref{bottomup}.
The algorithm initially constructs the sortable summarizations of all data series and sorts them using external sort.
Then it constructs in a bottom-up fashion a detailed iSAX index.
Finally this index is compacted by pushing more data series in the leaf nodes.

\begin{algorithm}[tb]
\small

\caption{Coconut-Trie: bottom-up bulk-loading of an prefix split based tree}
\label{bottomup}
\begin{algorithmic}[1]
\Procedure{Coconut-Trie}{\blue{rawFile}}
\While{not reached end of file}
   \State position = current file position;
   \State \blue{dataSeries= read data series of size n from rawFile;}
   \State SAX = convert dataSeries to SAX;
   \State invSAX = invertSum(SAX);
   \State Move file pointer n points;
   \State Add the (invSAX, position) pair to the buffer;
   \If{the main memory is full}
        \State Sort buffer according to invSAX
        \State Flush sorted buffer to the disk
   \EndIf
\EndWhile
\State Sort flushed runs using external sort

\While{not reached end of sorted file}
   \State Read the next (invSAX, position) in the buffer
   \If{the main memory is full}
        \For{every different subtree in buffer}
            \State //\textit{Move data from the buffer}
            \State //\textit{to leaf buffer}
            \State //\textit{and construct bottom-up the index}
            \For{every (invSAX, position) in buffer}
                 \State insertBottopUp(invSAX, position);
            \EndFor
            \State //\textit{merge leaf nodes as much as possible}
            \State CompactSubtree(root)
            \State //\textit{Flush all Leaf Buffers containing}
            \State //\textit{(Sax, position) pairs to the disk}
            \For{every leaf in subtree do}
                \State Flush the leaf to the disk;
            \EndFor
        \EndFor
  \EndIf
\EndWhile

\EndProcedure
\end{algorithmic}
\end{algorithm}

The input of the algorithm is a raw file, which contains all data series.
The process starts with a full scan of the raw data file in order to create the sortable summarizations for all data series (lines 4-6).
For data series we also record their offset in the raw file, so future queries can easily retrieve the raw values.
All sortable summarizations and offsets are stored in an FBL buffer (First Buffer Layer).
As soon as the buffer is full, it is sorted in the main memory and the sorted pairs are written to disk.

The process continues until we reach the end of the raw file.

If there are more than one sorted runs on disk, we sort them using external sort, and the final sorted file is written to disk.

Having the sortable summarizations sorted, all records that belong to a specific subtree are grouped together.
As such we exploit them in order to build a minimal tree in a bottom-up fashion, i.e., a tree that does not contain any raw data series (lines 22-24).
The main idea of the corresponding algorithm, i.e. the \textit{insertBottopUp} procedure, is that initially a new node is created for each different SAX representation.
Then, the algorithm replaces in iterations the least significant bits of the SAX representations with star marks until a common SAX prefix is identified to be placed in the parent node.
Then this idea is applied at the parent level and so on, until we reach the root (the corresponding algorithm is omitted due to lack of space).

The next phase is to compact this subtree, i.e. to push as many records in the leaf nodes as possible.
This is performed using the \textit{CompactSubtree} procedure (line 26).
To do that the algorithm iteratively checks whether the records of two sequential sibling nodes can fit together in a parent node.
If they do, the algorithm merges them and continues till all leaf nodes are visited.
Then the algorithm iterates again over the all leaves, until no more leaves are merged.
Finally each compacted subtree is flushed back to disk (lines 29-31).

The above algorithm is used to create a secondary index over the original raw file, keeping only the offsets in the leaf nodes. The algorithm performs the following passes over the data: (i) read the raw data series and compute the sortable summarizations; (ii) flush the sorted partitions of the summarizations to disk (along with their offsets); (iii) merge-sort them; and (iv) build the index.
This process involves $O(N/B)$ I/Os, but usually all the summarizations and their offsets fit in main memory, eliminating the need for passes (ii) and (iii).

A slight variation of the aforementioned algorithm could be used to create a fully-materialized iSAX index as well.\footnote{In a materialized index, the raw data-series are stored alongside their summarizations within the index, whereas in a non-materialized one the index contains \emph{pointers} to the raw data series that are stored in a different file.}
We call this variation \emph{Coconut-Trie-Full}.
This would require the raw data series to be sorted alongside their sortable summarizations in the sort-merge phase, and then flushed to disk.
Although the complexity of the algorithm would be the same, it would require additional passes in the sort-merge phase, and an additional pass over the raw data, in order to flush them to the leaf nodes.

\begin{figure}[tb]
    \includegraphics[width=\columnwidth]{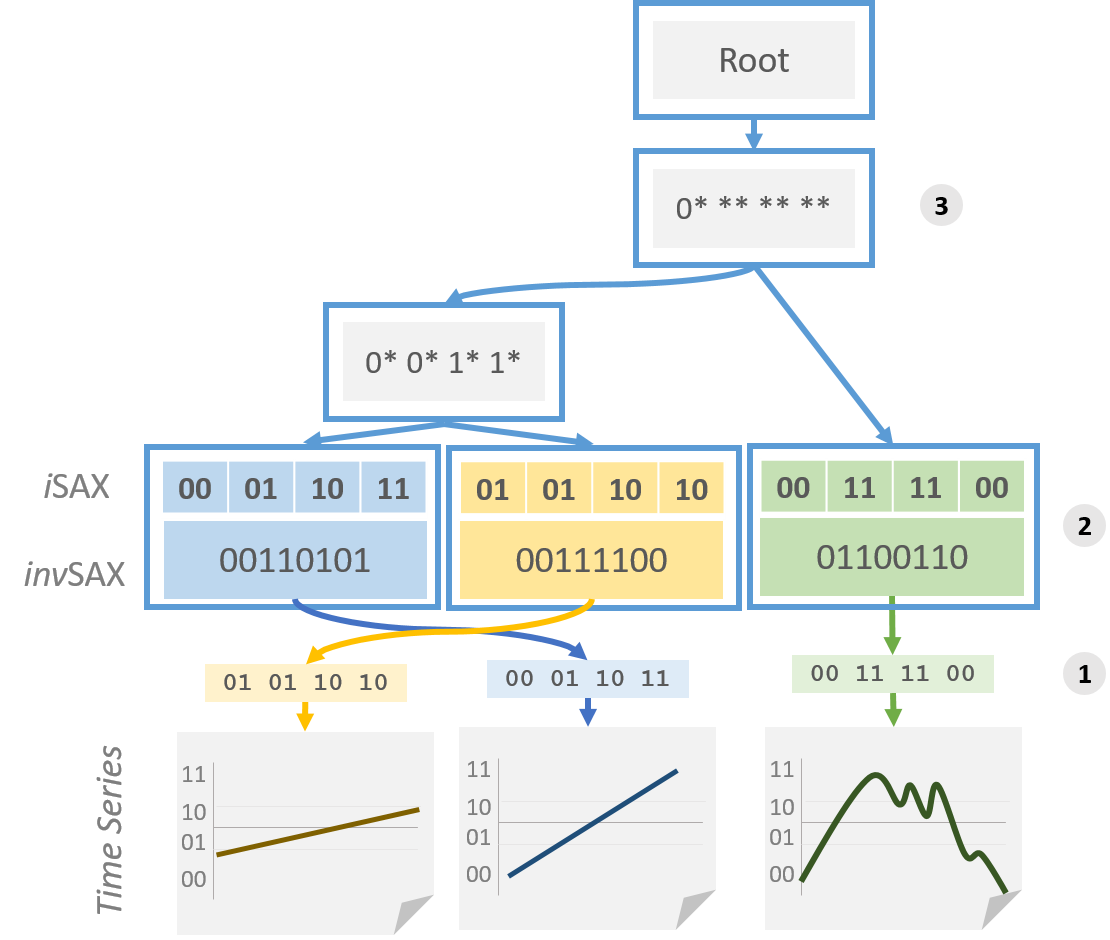}
    \caption{ Constructing bottom-up a Coconut-Trie index - before calling the \textit{compactSubtree} procedure. }
    \label{fig:example}
\end{figure}
\begin{exmp}
Figure~\ref{fig:example} illustrates an example of creating a Coconut-Trie index using the bottom-up Algorithm~\ref{bottomup}. As shown in the figure, we initially construct the summarizations (SAX) for all data series, as well as their sortable summarizations (invSAX).
Then, we sort them using their invSAX value, and we construct the corresponding Coconut-Trie index using the \textit{InsertBottomUp} algorithm.
Following this algorithm, initially, the first data series is placed in a new node.
The second data series is placed in yet a new node, since it has a different SAX representation than the first one.
Then, the \textit{createUptree} procedure is called to link the new node with the previous node. As such, the four least significant bits are replaced with stars, until the algorithm identifies a common prefix that could be used as the mask of the parent node $(0* 0* 1* 1*)$.
The parent is generated and linked to the root node.
The third data series is then inserted to the tree, and a new node is generated.
This node should be linked to the already existing tree: the \textit{createUptree} procedure is called again, using as input the SAX representations of the second and third data series.
The least significant bits are again replaced by a star, one by one until we identify the parent that should be generated linking the third node to the tree.
The resulting Coconut-Trie tree (refer to Figure~\ref{fig:example}) demonstrates the state of the tree before calling the \textit{CompactSubtree} procedure, which will follow in order to compact the entire tree.
Assuming that a leaf node can hold two data series, the corresponding algorithm will identify that the first two time-series have the same parent and they fit together.
As such they can be placed directly in their parent node, removing the child nodes.
\end{exmp}

\Paragraph{Queries} Since the constructed index is essentially no different than an iSAX index, we use the traditional approximate and exact search algorithms in order to perform querying.
Approximate search works by visiting the single most promising leaf, and calculating the minimum distance to the raw data series contained in it.
It provides answers of good quality (returns a top 100 answer for the nearest neighbor search in 91.5\% of the cases for iSAX with extremely fast response times~\cite{Shieh2008}).
On the other hand, exact search guarantees that we get the exact answer, but with potentially much higher execution time.
For exact search, we employ the SIMS algorithm, implementing a skip sequential scan algorithm, shown to outperform traditional exact search algorithms~\cite{ZoumpatianosIP16}.

\subsection{Coconut-Tree}
\label{sec:tree}

Although Coconut-Trie achieves contiguity, i.e. adjacent leaf nodes are placed next to each other in storage, a lot of disk space is wasted in those leafs: many of them are half-full or less, due to the way the index is constructed (i.e., compacting child nodes to a parent one).
In addition, since the constructed tree in both Coconut-Trie and in current state-of-the-art are unbalanced trees, they offer no guarantees for the query answering time.

We now present Coconut-Tree, a data series index that organizes data series based on sortable summarizations, and improves upon Coconut-Trie by eliminating the constraint that a node can only contain elements with a common prefix.
This leads to a balanced index that can densely pack data in its leaf nodes (at a fill-factor that can be controlled by the user).
The corresponding algorithm completes index construction again in $O(N/B)$ time.

Index construction, shown in Algorithm~\ref{ubtree}, receives the raw data file as input. 
A buffer is initialized, and while the buffer is not full the next data series is loaded from the raw file, and the sortable summarization is calculated and stored along with the position of this data series in the raw data file (lines 2-8).
Whenever the buffer fills up, it gets sorted and flushed to storage as an independent sorted partition (line 9-13). Ultimately, all sorted partitions get sort-merged into a single sorted partition (line 14). Some padding may be left in each storage block as space for future insertions. Internal nodes are then built on top of this sorted partition to construct a B-tree (line 15). 

\begin{algorithm}[tb]
	\caption{Coconut-Tree: Bottom-up bulk-loading of a balanced tree}
	\label{ubtree}
	\begin{algorithmic}[1]
		\small
		\Procedure{Coconut-Tree}{\blue{rawFile}}
		\While{not reached end of file}
		\State position = current file position;
		\State \blue{dataSeries = read $n$ data series from rawFile;}
		\State iSAX = convert dataSeries to iSAX;
		\State invSAX = invertSum(iSAX);
		\State Move file pointer n points;
		\State Add the (invSAX, position) pair in the buffer;
		\If{the main memory is full}
		\State Sort buffer according to invSAX
		\State Flush sorted buffer to the disk
		\EndIf
		\EndWhile
		\State Merge-sort all flushed runs 
		\State Build internal nodes on top of sorted file
		\EndProcedure
	\end{algorithmic}
\end{algorithm}

The Algorithm~\ref{ubtree} builds a secondary index with only offsets in the lead nodes, but it can be used to construct a fully materialized index as well, where all data reside in the leaf nodes.
We call the materialized version of the algorithm \emph{Coconut-Tree-Full}.
We expect that index construction time of Coconut-Tree-Full will be significantly larger.
Nevertheless, we also expect that query execution time would be better, since it will not perform additional I/Os to go to the raw data file for accessing each required data series record.

\begin{figure}[tb]
	\centering
    \includegraphics[width=\columnwidth]{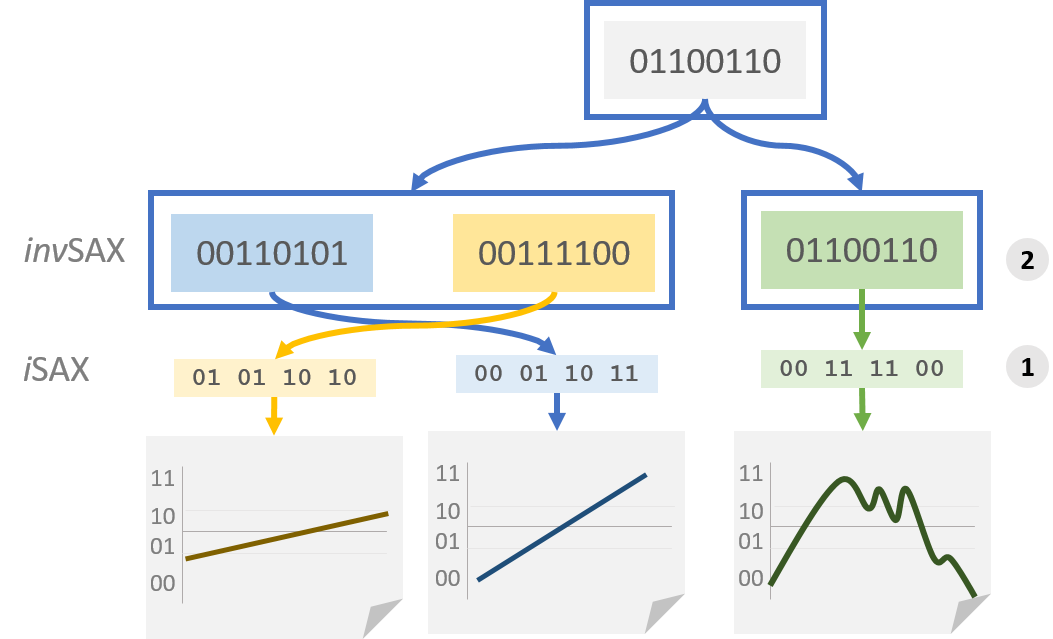}
    \caption{ Constructing a Coconut-Tree index. }
    \label{fig:examplecoconut}
\end{figure}

\begin{exmp}
Figure~\ref{fig:examplecoconut} illustrates the construction of a Coconut-Tree index.
Initially, we construct for all data series their SAX and their invSAX representations.
We then sort them using their invSAX value, and we construct the Coconut-Tree index in a bottom-up fashion (exploiting the bulk-loading algorithm for UB-Trees~\cite{DBLP:books/daglib/0011128}).
Note that the constructed index in this case is balanced.
\end{exmp}

\Paragraph{Querying} For \textit{approximate search}, when a query arrives (in the form of a data series), it is first converted to its sortable summarization.
Then the Coconut-Tree index is traversed searching for this sortable summarization similar the approximate search in iSAX trees.
The idea is to search for the leaf, where the query series would reside if it was part of the indexed data set.
If such a record exists, it is retrieved from the disk and returned to the user.
On the other hand, if such a record does not exist, all data series in a specific radius from this specific point are retrieved from the disk (usually a disk page), and their real distances from the query are calculated.
The data series with the minimum distance found among the data series in that radius is used as the approximate answer.
Thus, in terms of execution cost, the algorithm visits as many nodes as the depth of the tree, and any additional leaf nodes within the selected radius.

Note that in a Coconut-Tree index, we have pointers between neighboring leaves, which are allocated sequentially on disk.
This allowed us to experiment with the radius size, optimizing the trade-off between the quality of the answer and the execution time of the approximate search.

 \begin{algorithm}[tb]
    \caption{Approximate search for the Coconut-Tree}
    \label{approxsearch}
    \begin{algorithmic}[1]
    	\small
\Procedure{approxSearchCoconutTree}{dataSeries, invSAX, index, radius}
        \State targetPoint = point where invSAX should be inserted
        \State //\textit{Calculate the real leaf distance between }
        \State //\textit{the dataSeries and the raw data series }
        \State //\textit{in a radius around the place that the }
        \State //\textit{dataSeries should reside if existed}
        \State bsf = caclRadLeafDist(targetPoint, dataSeries, radius);
\EndProcedure
    \end{algorithmic}
\end{algorithm}

For implementing \textit{exact search} for Coconut-Tree, we implement a skip sequential scan algorithm (refer to Algorithm~\ref{exactsearch}) similar to SIMS~\cite{ZoumpatianosIP16}.
Our algorithm employs approximate search as a first step in order to prune the search space.
It then accesses the data in a sequential manner, and finally produces an exact, correct answer. We call this algorithm Coconut-Tree Scan of In-Memory Summarizations (CoconutTreeSIMS).
The main intuition is that while the raw data do not fit in main memory, their summarized representations (which are orders of magnitude smaller) will fit in main memory (remember that the SAX summaries of 1 billion data series occupy merely 16 GB in main memory). By keeping these data in-memory and scanning them, we can estimate a bound for every data series in the data set.

 \begin{algorithm}[tb]
    \caption{Coconut-Tree Scan of In-Memory summarizations }
    \label{exactsearch}
    \begin{algorithmic}[1]
    	\small
\Procedure{coconutTreeSIMS}{dataSeries, invSAX, index, radius}
        \State //\textit{if SAX sums are not in memory, load them}
        \If{invSums = 0}
          \State invSums = loadinvSaxFromDisk();
        \EndIf
        \State //\textit{perform an approximate search}
        \State bsf = approxSearchCoconutTree(dataSeries, invSAX, index, radius);
        \State //\textit{Compute minimum distances for all summaries}
        \State Initialize mindists[] array;
        \State //\textit{use multiple threads \& compute bounds in parallel}
        \State parallelMinDists(mindists, invSums, dataSeries);
        \State //\textit{Read raw data for unprunable recorde}
        \State recordPosition = 0;
         \For{every mindist in mindists}
            \If{mindist < bsf}
                  \State rawData = read raw data series from index;
                  \State realDist = Dist(rawData, dataSeries);
                  \If{realDist < bsf}
                    \State bsf = realDist;
                  \EndIf
            \EndIf
            \State recordPosition++;
        \EndFor

\EndProcedure
    \end{algorithmic}
\end{algorithm}

The algorithm differs from the original SIMS algorithm in that it searches over the sorted invSAX representations for the initial pruning, and it then uses the Coconut-Tree index to get the raw data-series instead of accessing the original file with the raw data series.
As such, Algorithm~\ref{exactsearch} starts by checking whether the sortable summarization data are in memory (lines 3-4), and if not it loads them in order to avoid recalculating them for each query.
It then creates an initial best-so-far (bsf) answer (line 7), using the approximate search algorithm described previously (Algorithm~\ref{approxsearch}).
A minimum distance estimation is calculated between the query and each in-memory sortable summarization (line 11) using multiple parallel threads, operating on different data subsets.
For each lower bound distance estimation, if this is smaller than the real distance to the bsf, we fetch the complete data series from the Coconut-Tree index, and calculate the real distance (lines 15-22).
If the real distance is smaller than the bsf, we update the bsf value (lines 19-21).
Since the summaries array is aligned to the data on disk, what we essentially do is a synchronized skip sequential scan of the raw data and the in-memory mindists array.
This property allows us to prune a large amount of data, while ensuring that the executed operations are very efficient: we do sequential reads in both main memory and on disk, and we use modern multi-core CPUs to operate in parallel on the data stored in main memory.
At the end, the algorithm returns the final bsf to the user, which is the exact query answer.


\subsection{Coconut-LSM}
\label{sec:lsm}

While Coconut-Tree creates a compact and contiguous index that can be constructed and queried efficiently, it does not perform well in the presence of random insertions (i.e., that are uniformly distributed across the key space). The reason is that when insertions are randomly distributed, each of them requires $O(1)$ I/O to process (i.e., one I/O to read the corresponding node and another I/O to rewrite it). For insertion-heavy workloads, this can harm throughput. To mitigate this problem, we introduce Coconut-LSM, a new write-optimized data series index based on sortable summarization.  

Coconut-LSM organizes the data series summarizations as an LSM-tree \cite{DBLP:journals/acta/ONeilCGO96,dayan2017}. The core idea is to buffer incoming insertions in memory, to flush the buffer to storage as an independent sorted run every time that it fills up, and to bound the overall number of runs in storage by gradually sort-merging them to restrict read cost (i.e., the number of runs a read has to search). LSM-tree sort-merges runs of similar sizes, and it organizes them into levels of exponentially increasing capacities. We use a variation of LSM-tree with a size ratio of 2 between the capacities of every pair of adjacent levels. As a result, there are at most {\small $O(log_2(N))$} runs in the system. 
Since every insertion gets merged across each level, and since every I/O during sort-merge copies $B$ entries, the amortized cost per insertion is {\small$O(\frac{log_2(N)}{B})$} I/O. Since the storage block size $B$ is large, the insertion cost for Coconut-LSM gets amortized and is therefore significantly lower than for any existing data series index. Thus, Coconut-LSM enables more efficient insertions at the expense of slightly more expensive queries. 


The construction algorithm is similar to the one for Coconut-Tree, performing a two-pass external sort of the data in {\small$O(N/B)$} I/O, and is shown in Algorithm~\ref{lsmtree}. The resulting sorted file, also called a run, becomes the largest level of the LSM-tree. 
Similarly to Coconut-Tree, we also consider a materialized variant of Coconut-LSM called \emph{Coconut-LSM-Full}, which stores raw data series within the LSM-tree, and which we evaluate later. 



\begin{algorithm}[tb]
	\caption{Coconut-LSM: Bottom-up bulk-loading of an LSM-tree}
	\label{lsmtree}
	\begin{algorithmic}[1]
		\small
		\Procedure{Coconut-LSM}{rawFile}
		\While{not reached end of file}
		\State position = current file position;
		\State dataSeries = read $n$ data series from rawFile;
		\State iSAX = convert dataSeries to iSAX;
		\State invSAX = invertSum(iSAX);
		\State Move file pointer n points;
		\State Add the (invSAX, position) pair in the buffer;
		\If{the main memory is full}
		\State Sort buffer according to invSAX
		\State Flush sorted buffer to the disk
		\EndIf
		\EndWhile
		\State Sort flushed runs using external sort
		\State Use LSM-Tree bulk-loading algorithm to build a tree on top of the sorted file and record the individual flushes on disk
		\EndProcedure
	\end{algorithmic}
\end{algorithm}

\begin{figure*}[tb]
	\centering
	\includegraphics[width=0.9\textwidth]{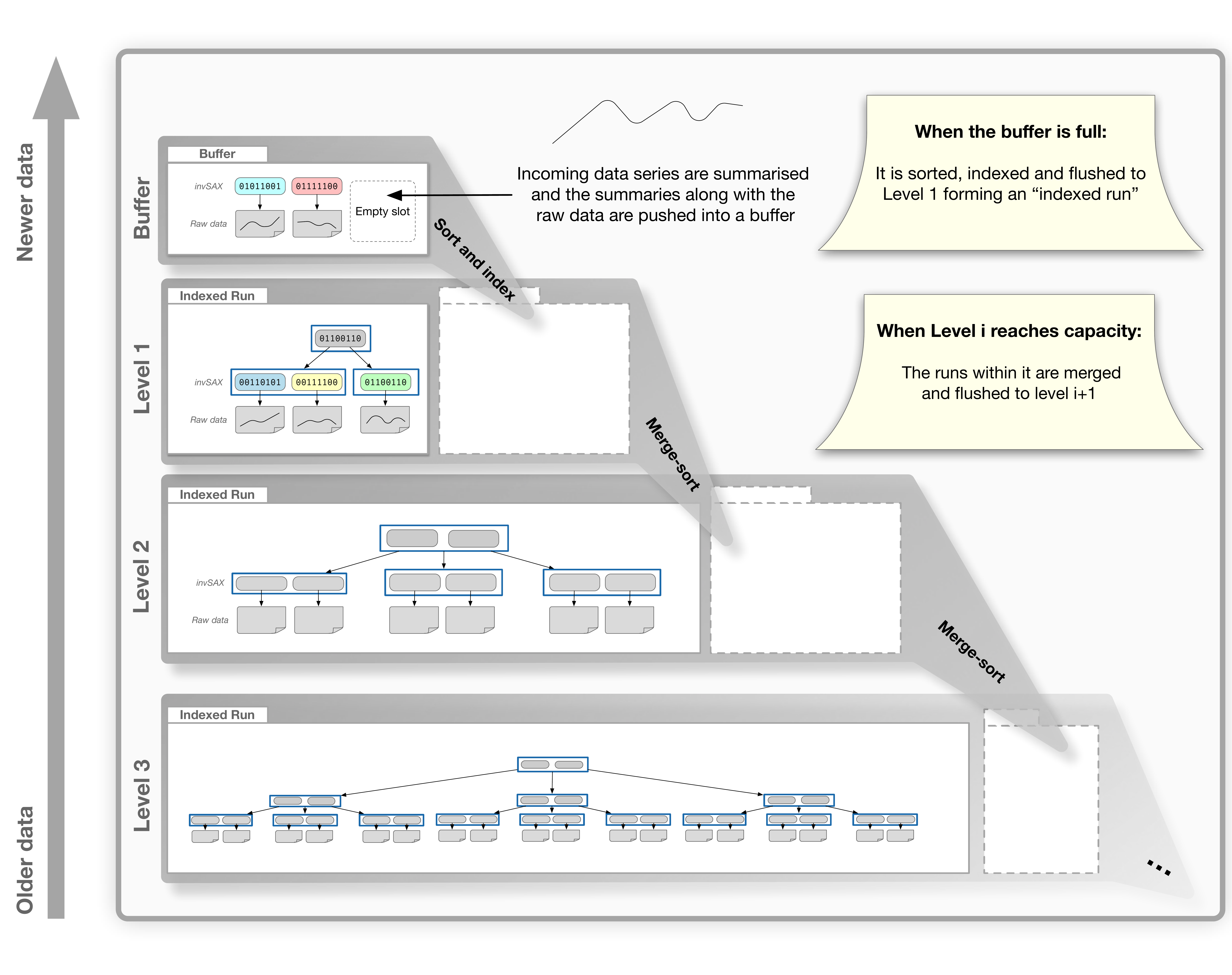}
	\caption{ Constructing a Coconut-LSM index.}
	\label{fig:examplecoconutlsm}
\end{figure*}

\begin{exmp}
Figure~\ref{fig:examplecoconutlsm} illustrates the construction of a Coconut-LSM index.
Initially, we construct for all data series their invSAX representations. We then sort them using their invSAX value, and we construct the Coconut-LSM index in a bottom-up fashion (exploiting the bulk-loading algorithm for LSM-Trees). The bulk loading algorithm buffers incoming insertions in memory and flushes the buffer to storage as it fill-up, creating multiple Coconut indexes. 
 As multiple indexes are constructed in the incoming buffer (also referred to as level 0), they are asynchronously merged to form larger indexes in level 1. The same applies for level 1 indexes that are asynchronously merged to formulate larger, level 2 indexes.
\end{exmp}

\Paragraph{Querying} For \textit{approximate search}, Algorithm \ref{approxsearch} for Coconut-Tree is applied to each individual run of Coconut-LSM. 
The data series with the minimum distance found across the runs of Coconut-LSM is used as the approximate answer. 
Note that approximate search in Coconut-LSM is more expensive in terms of I/Os than Coconut-Tree, as multiple runs need to be searched. 
In this way, Coconut-LSM trades approximate query performance for insertion performance. 

For implementing \textit{exact search} for Coconut-LSM, we revisit the corresponding algorithm for Coconut-Tree. The new algorithm is shown in Algorithm~\ref{exactsearchLSM}. In the first step, the algorithm employs approximate search in order to prune the search space. It then accesses the subtrees in a sequential manner, and finally produces an exact, correct answer. We call this algorithm Coconut-LSM Scan of In-Memory Summarizations (Coconut-LSM-SIMS).

The main intuition for this algorithm is that we would like to search sequentially all subtrees of the LSM tree in order to optimize read, still however performing a skip sequential scan. As such we use the snapshot of the available summarizations produced in indexing phase. By keeping these data in-memory and scanning them, we can estimate a bound for every data series in the data set.

\begin{algorithm}[tb]
	\caption{Coconut-LSM Scan of In-Memory summarizations }
	\label{exactsearchLSM}
	\begin{algorithmic}[1]
		\small
		\Procedure{coconut-LSM-SIMS}{dataSeries, invSAX, index, radius}
		\State //\textit{if SAX sums are not in memory, load them}
		\If{invSums = 0}
		\State invSums = loadinvSaxFromDisk();
		\EndIf
		\State //\textit{perform an approximate search}
		\State bsf = approxSearchCoconutTree(dataSeries, invSAX, index, radius);
		
		\For{every subtree of the LSM structure}
		\State //\textit{Compute minimum distances for all summaries}
		\State Initialize mindists[] array;
		\State //\textit{use multiple threads \& compute bounds in parallel}
		\State parallelMinDists(mindists, invSums, dataSeries);
		\State //\textit{Read raw data for unprunable recorde}
		\State recordPosition = 0;
		\For{every mindist in mindists}
		\If{mindist < bsf}
		\State rawData = read raw data series from index;
		\State realDist = Dist(rawData, dataSeries);
		\If{realDist < bsf}
		\State bsf = realDist;
		\EndIf
		\EndIf
		\State recordPosition++;
		\EndFor
		\EndFor
		
		\EndProcedure
	\end{algorithmic}
\end{algorithm}

\section{Sliding windows}
\label{sec:sliding}

Up until now, we focused on nearest neighbor search across an entire dataset. 
In many modern applications, however, queries have temporal constraints: they must find the nearest neighbor from within the most recent data (e.g., in infrastructure monitoring, or geo-temporal applications). 
The size of the temporal window of interest often varies across and within applications to enable different granularities of analysis (e.g., data from the past week, month, year, etc.). Therefore, a data series index needs to flexibly support variable-sized window queries. Ideally, it should save on storage bandwidth by avoiding access to data that is older than a specified query window. 

In this section, we describe three approaches for supporting window queries. The first two approaches, \textit{post-processing} and \textit{temporal partitioning}, only support efficient long or short window queries, respectively, but neither supports both. These two approaches represent the best we can do with existing data series indexes as well as with Coconut-Trie and Coconut-Tree. We then show how Coconut-LSM enable a third approach that supports window queries of any size efficiently. We coin it \textit{bounded temporal partitioning} (BTP).
For all three approaches, we attach a timestamp to each entry. We experimentally compare them in Section \ref{sec:eval}. 
	

\subsection{Approach 1: Post-Processing (PP)} 
Post-processing relies on examining the timestamp of every entry as it is encountered during query processing and discarding it if the timestamp does not fit within the window specified by the query. 
Exact queries take place as before, with the difference that they now also check every entry's timestamp. Approximate queries, however, may need to broaden the scope of their search if the first node that is encountered only contains entries that are outside of the specified window. Hence, we adapt them to incrementally expand their search across adjacent leaf nodes until an entry within the specified window is found. 


While post-processing is the simplest approach to implement, it is inefficient for exact queries if the specified time window encompasses a small proportion of the data. The reason is that it does not allow to save storage bandwidth by  avoiding access to older entries. Hence, an exact query to the most recent data consumes as much storage bandwidth as a query over the entire data. For approximate queries, search may also take significantly longer to execute as potentially many nodes need to be searched until an entry within the window is found. 

\subsection{Approach 2: Temporal Partitioning (TP)}

With temporal partitioning, we create a new index partition based on the in-memory buffer's contents every time that the buffer fills up. In this way, the system gathers more and more temporal partitions over time, and it organizes them based on their creation time. This allows both approximate and exact queries to access only indexes whose creation timestamp falls within or intersects with a specified query window. 

TP works well for short window queries as it allows them to skip access to most of the data in storage. On the other hand, it performs poorly for windows that span a significant proportion of the data. For exact queries, the reason is that they must begin the search from scratch for every partition, and so they cannot leverage the lower-bounding property of invSAX as effectively to spatially prune within each partition. For approximate queries, the reason is that they need to issue one I/O to every qualifying partition (potentially hundreds for large data sizes).

\subsection{Approach 3: Bounded Temporal Partitioning (BTP)}
While neither the first nor the second approach supports both long and short window queries efficiently, many applications need to be able to use both short and long window sizes, while maintaining good performance in all cases. Our insight is that Coconut-LSM enables a new approach that combines the best of these two approaches. 
By design, Coconut-LSM creates a new temporal partition every time the buffer flushes (as with TP), and it sort-merges temporal partitions of similar sizes. In this way, newer data resides in smaller partitions, while older data gradually moves to larger contiguous partitions. This allows queries over short windows to save storage bandwidth by skipping larger partitions. At the same time, it allows exact queries over long windows to spatially prune a greater proportion of the data by leveraging the lower bounding property of invSAX more effectively, and it allows approximate queries over long windows to issue fewer I/Os by bounding the overall number of partitions that need to be accessed.

We refer to this windowing approach as Bounded Temporal Partitioning (BTP). We implement BTP on top of Coconut-LSM by modifying it to take a window size as a query parameter, and to skip accessing larger partitions that fall outside of a specified window size. We demonstrate the benefits of BTP for both small and large window queries in Section \ref{sec:eval}.


Note that with unsortable summarizations (as is the case with the traditional state-of-the-art data series indexes), the BTP approach would have been inviable, as it would have to rely on expensive in-place insertions for merging partitions. We therefore observe here, too, that the ability to sort the summarizations opens up new opportunities for optimization that would have been impossible otherwise.

\begin{figure*}[tb]
	\centering
	\includegraphics[width=\textwidth]{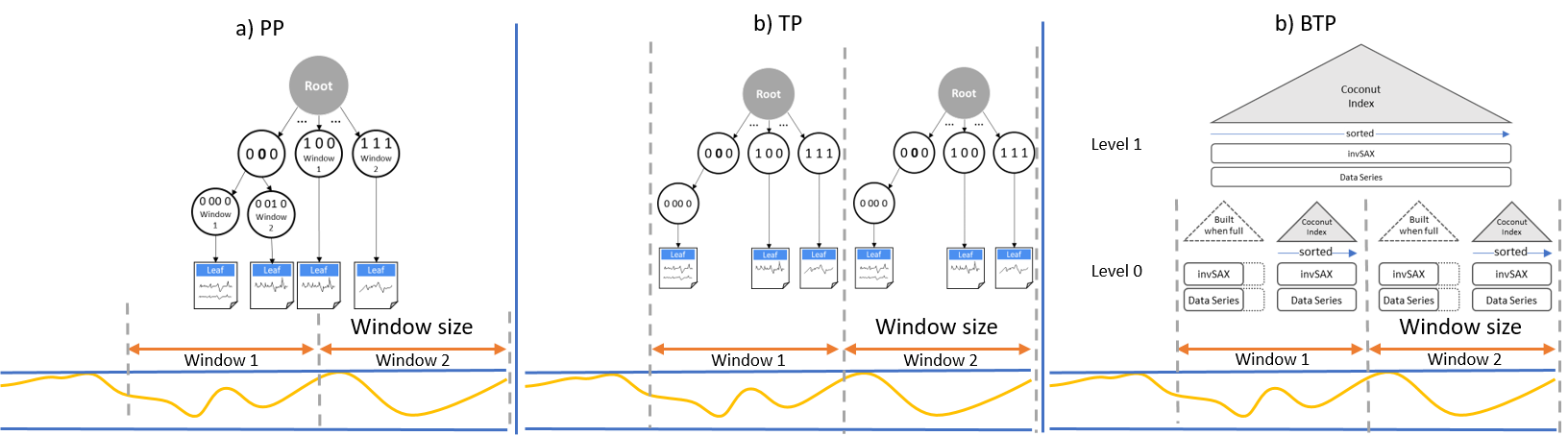}
	\caption{The various sliding windows appoaches: (a) Post-Processing (PP), (b) Temporal Partitioning (TP) and (c) Bounded Temporal Partitioning (BTP).}
	\label{fig:slidingwindowst}
\end{figure*}

\begin{exmp}
	
	Figure~\ref{fig:slidingwindowst} illustrates schematically the three approaches. In the first case, PP, a full index is constructed that covers the entire dataset.
	For TP, multiple indexes are constructed, each one for a different window partition of the data. 
	Finally, in the BTP approach, the index is constructed containing all entries, however creating a temporal partition each time the buffer is flushed to disk. Therefore, it guarantees optimal access to window queries, but it also enables querying records that reside on other sliding window sizes. 
\end{exmp}

\section{Experimental Evaluation} \label{sec:eval}

In this section, we present our experimental evaluation.
We demonstrate the benefits of sortability, enabling a variety of new choices for data structures to be used for better space-efficiency and for more efficiently bulk-loading, querying, and updating the data.

\Paragraph{Algorithms}
We benchmark all indexing methods presented in this paper against the state-of-the-art data series indexing techniques. 
More specifically, we compare our materialized methods with R-tree~\cite{Guttman1984},  Vertical~\cite{DBLP:conf/kdd/KashyapK11}, DSTree~\cite{Wang2013} and ADS-Full~\cite{ZoumpatianosIP16}, and our non-materialized methods with ADS+~\cite{ZoumpatianosIP16} and a non-materialized version we implemented over R-tree, the R-tree+.

The Vertical approach generates an index using data series features, obtained by a multi-resolution Discrete Wavelet Transform, in a stepwise sequential-scan manner, one level of resolution at a time.
DSTree is a data adaptive and dynamic segmentation tree index that provides tight upper and lower bounds on distances between time series.
ADS-Full is an algorithm that constructs an $i$SAX clustered index by performing two passes over the raw data series file.
ADS+ is an adaptive data structure, which starts by building a minimal secondary index.
Leaf sizes are refined during query answering, and leaves are materialized on-the-fly. As such query answering has the additional overhead of the refinement of the leafs.
The R-tree index is built on the raw data series by indexing their PAA summarizations.
The raw data series are stored in the leaves of the tree.
Our R-tree implementation uses the Sort-Tail-Recursive bulk loading algorithm~\cite{leutenegger1997}.
R-tree+ is the non-materialized version of the R-tree, using file pointers in the leaves instead of the original time series.
In our experiments, we used the same leaf size (2000 records) for all indexing structures.

In the experiments on index construction and querying, we do not include Coconut-LSM. The reason is that in the absence of insertions, Coconut-LSM after bulk-loading contains all of its data in one level, and a one-level LSM-tree is structurally equivalent to a B-tree \cite{DBLP:journals/acta/ONeilCGO96}. We therefore include Coconut-LSM in the experiments when we also introduce insertions into the workloads. 

\Paragraph{Infrastructure}
All algorithms are compiled with GCC 4.6.3 under Ubuntu Linux 12.04 LTS. We used an Intel Xeon machine with
5x2TB SATA 7.2 RPM hard drives in RAID 0.
The memory made available for each algorithm was controlled according to the experiment.

\Paragraph{Datasets}
\begin{figure}[tb]
	\begin{center}		
		\includegraphics[width=\columnwidth]{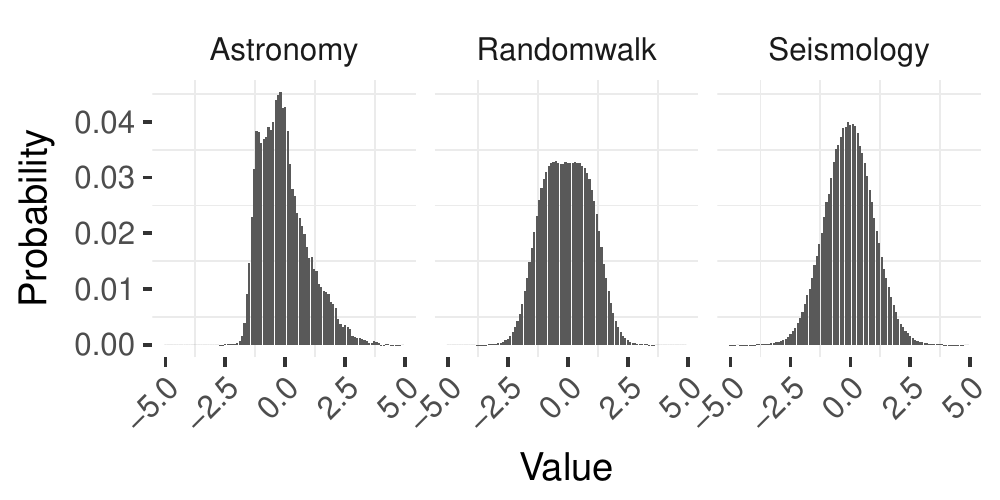}
		\caption{Value histograms for all datasets used.}
		\label{fig:distributions}
	\end{center}
\end{figure}
For our experiments we used both synthetic and real datasets. Synthetic datasets were generated using a random walk data series generator: a random number is drawn from a Gaussian distribution (0,1); then, at each time point a new number is drawn from this distribution and added to the value of the last number.
This kind of data has been extensively used in the past (see~\cite{ZoumpatianosLPG15} for a list of references), and has been shown to effectively simulate real-world financial data~\cite{DBLP:conf/sigmod/FaloutsosRM94}.

The real datasets we used in our experiments are seismic and astronomy data.
We used the IRIS Seismic Data Access repository~\cite{iris} to gather data series representing seismic waves from various locations.
We obtained 100 million data series by extracting one sample per second from the original data series, and then partitioning them into smaller series of 256 samples each by sliding every 4 samples over the original series. 
The complete dataset size was 100GB.
For the second real dataset, we used astronomy data series representing celestial objects~\cite{refId0}.
The dataset comprised of 270 million data series, obtained by partitioning the original series into smaller series of 256 samples each using a sliding step of one sample. The total dataset size was 277GB.

All our datasets have been z-normalized by subtracting the mean and dividing by the standard deviation.
This is a requirement by many applications that need to measure similarity irrespective of translation and scaling of the data series~\cite{DBLP:conf/cp/GoldinK95}.
Moreover, it allows us to compute correlations based on the euclidean distance values~\cite{MueenNL10}.

In Figure~\ref{fig:distributions}, we show the distributions of the values for all datasets.
The distributions of the synthetic and seismology data are very similar, while astronomy data are slightly skewed.

\Paragraph{Query Workloads}
Each query is given in the form of a randomly selected data series $q$ and having the index try to locate whether this data series or a similar one exists in the database.
For querying the real datasets we obtained additional data series from the raw datasets using the same technique for collecting the datasets to be used in the query workload.

\begin{figure}[tb]
	\begin{center}		
		\includegraphics[width=\columnwidth]{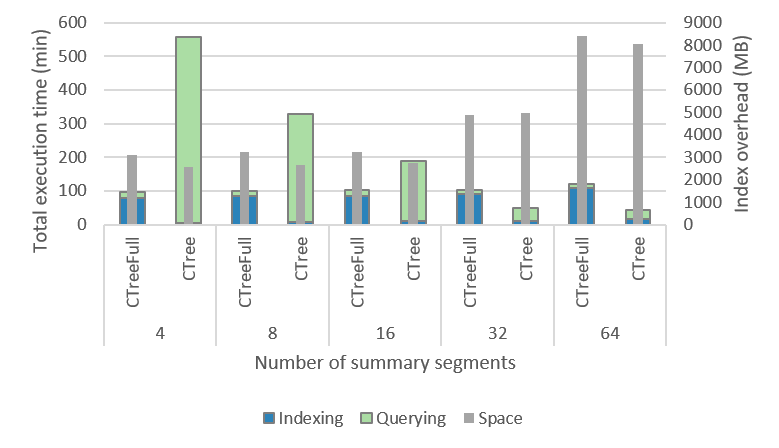}
		\caption{Indexing \& Querying data series using variable number of summary segments.}
		\label{fig:variablesegments2}
	\end{center}
\end{figure}

\Paragraph{Configuring number of segments}
As a first step, before comparing with other approaches, we studied the effect of the number of segments of the generated summaries on performance. The idea was to evaluate the trade-off between number of segments, space overhead introduced by the indexing structure over the raw data, and indexing and querying execution time. 
We used a synthetic data series collection of 100GB data series and 100 exact queries, using limited memory (100K data series) for both indexing and querying. The accumulative execution time for both querying and indexing is shown in Figure~\ref{fig:variablesegments2}, where we can also see the index space overhead in each case (thin gray line). 
As shown, the larger the number of segments, the larger the indexing time for both materialized (CTreeFull) and non-materialized (CTree) approaches. In addition, the benefit in indexing has an impact on querying, as smaller summaries cannot prune effectively the search space when performing exact queries. On the other hand going beyond 16 number of segments almost doubles the additional space introduced by our indexing structures. 
Therefore, we selected 16 as the number of segments. 
Unless mentioned otherwise, in the rest of the experimental evaluation, the summarizations use 16 SAX words, the size of data series was of 256 points, and each point has a floating precision of 4 bytes.

\subsection{Indexing}

\begin{figure*}[tb]
	\begin{subfigure}{0.32\textwidth}
		\begin{center}
			\includegraphics[width=\textwidth]{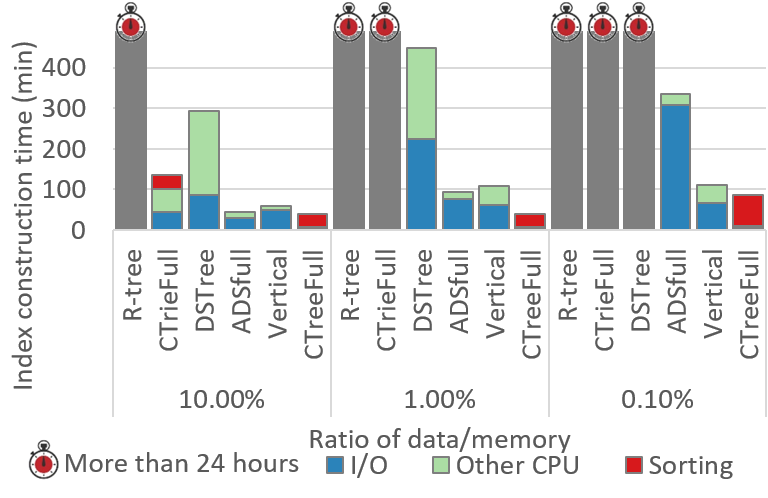}
			\caption{Index construction - materialized.}
			\label{fig:fixdata}
		\end{center}
	\end{subfigure}
	\begin{subfigure}{0.32\textwidth}
		\begin{center}
			\includegraphics[width=\textwidth]{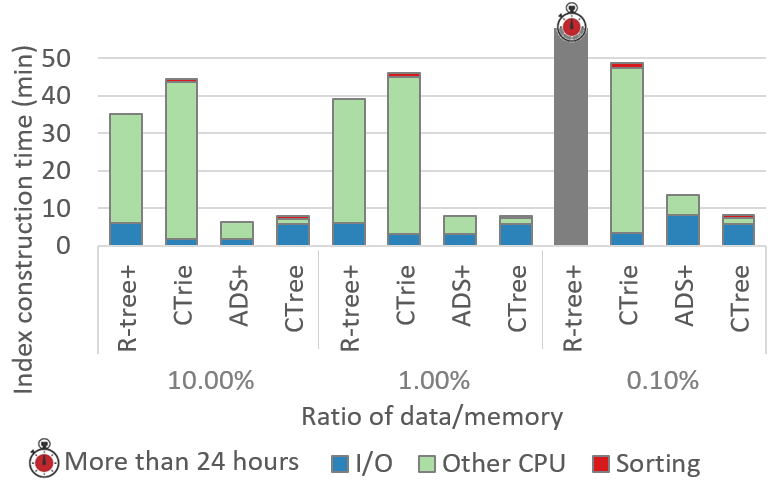}
			\caption{Index construction - non-materialized.}
			\label{fig:fixdata+}
		\end{center}
	\end{subfigure}
\begin{subfigure}{0.32\textwidth}
	\begin{center}
		\includegraphics[width=\textwidth]{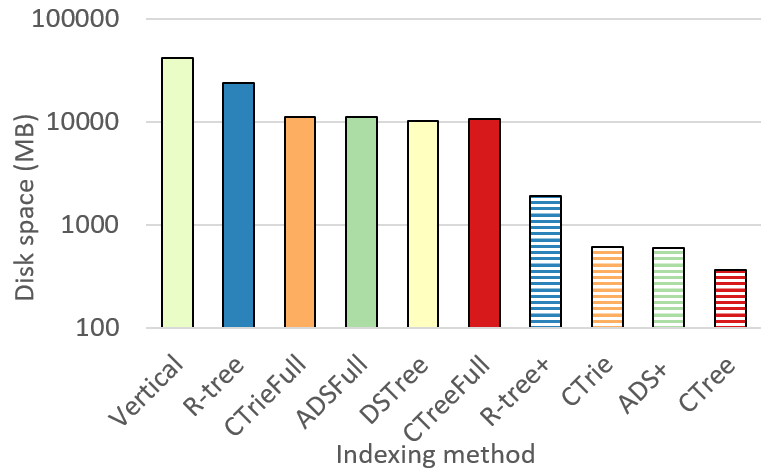}
		\caption{Indexing space overhead.}
		\label{fig:space}
	\end{center}
\end{subfigure}
\begin{subfigure}{0.32\textwidth}
	\begin{center}
		\includegraphics[width=\textwidth]{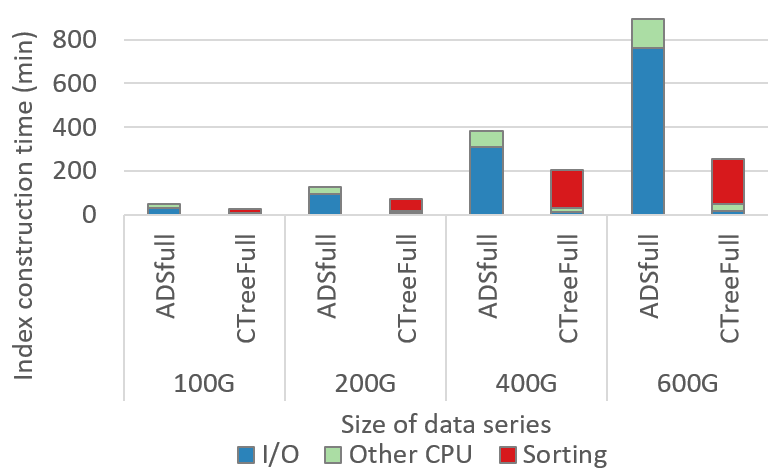}
		\caption{Index construction - materialized.}
		\label{fig:fixmemory}
	\end{center}
\end{subfigure}
\begin{subfigure}{0.32\textwidth}
	\begin{center}
		\includegraphics[width=\textwidth]{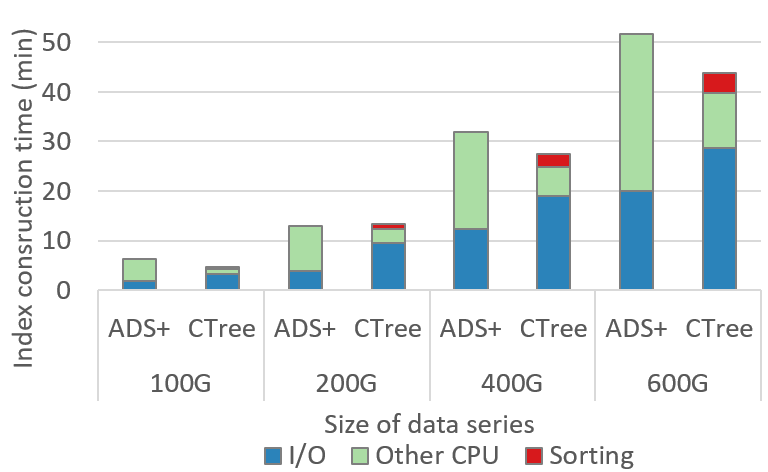}
		\caption{Index construction - non-materialized.}
		\label{fig:fixmemory+}
	\end{center}
\end{subfigure}
	\begin{subfigure}{0.32\textwidth}
	\begin{center}
		\includegraphics[width=\textwidth]{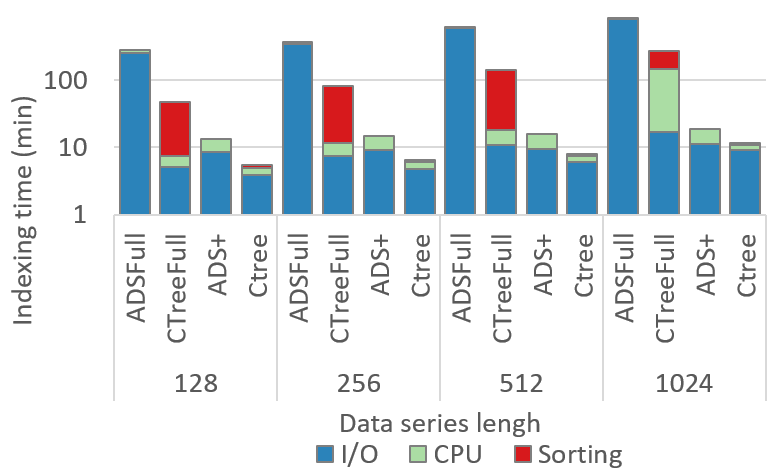}
		\caption{Indexing data series of different lengths.}
		\label{fig:variable}
	\end{center}
\end{subfigure}

	\caption{Indexing.}
\end{figure*}

\begin{figure}[tb]
	\begin{center}		
		\includegraphics[width=\columnwidth]{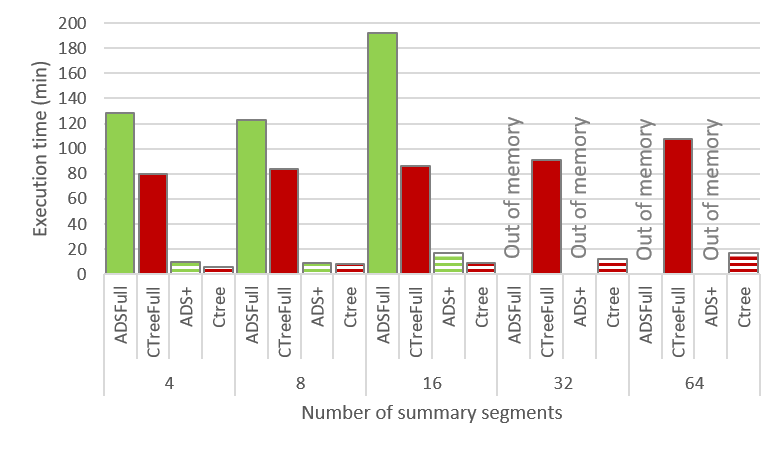}
		\caption{Indexing data series different number of summary segments.}
		\label{fig:variablesegments}
	\end{center}
\end{figure}


In our first set of experiments, we evaluate index construction speed. The results for the materialized algorithms are shown in Figure~\ref{fig:fixdata} as we vary the memory budget for each method to control the amount of buffering and caching they are able to leverage. We observe that Coconut-Tree-Full (CTreeFull) exhibits the best construction speed in all cases it is able to externally sort the raw data file.
As memory becomes limited, external sorting degrades gracefully in terms of performance. 
The construction time of Coconut-Trie-Full (CTrieFull) on the other hand, significantly increases as we constrain the memory (and the corresponding buffering), due to the extensive I/Os spent on the last pass of the data, for loading the unsorted raw data to the sorted leaves.
Moreover, we observe that Vertical is slower in all cases, 
while R-tree performs rather poorly.
The STR algorithm~\cite{leutenegger1997} that R-tree uses first sorts based on the first dimension into $N^{\frac{1}{D}}$ slabs (where $N$ is the number of points in a $D$-dimensional space), and then recursively repeats the process within each slab with one less dimension. 
As a result, runtime is the product of the number of elements and the number of dimensions: $O(N \cdot D)$ I/Os. 
In contrast, our implementation uses sortable summarizations to sort based on all dimensions with just one pass, amounting to $O(N)$ I/Os.
Finally, DSTree requires more than 24 hours to finish in most of the cases, as it inserts all data series in the index one by one, in a top-down fashion.
This requires multiple iterations to be performed over the raw data during splits in order to create more detailed summarizations, leading to a high I/O overhead.

In the non-materialized versions of the algorithms, shown in Figure~\ref{fig:fixdata+}, ADS+ is slightly better than Coconut-Tree (6.3 vs 7.8 mins), when given ample memory.
However when we restrict the available main memory, Coconut-Tree becomes faster than ADS+ (8.2 vs 13.4 mins).
This is due to the fact that as the leaves in ADS+ split, they cause random disk I/Os.
This slows down index construction, since buffering is limited when the main memory is limited.
On the other hand, Coconut-Trie (CTrie) spends a significant time in compacting its nodes, which significantly slows down index construction.
The performance of R-tree+ matches the behavior of the materialized R-tree, requiring much more execution time than the leading approaches.

Finally, we observe that non-materialized versions outperform the materialized ones, since they do not store the entire dataset, but only the summarizations and pointers to the raw data file.
Moreover, we note that sorting in the non-materialized versions is really fast, since only the summarizations need to be sorted, and so far less data has to be moved and reshuffled.

\Paragraph{Space} Since storage space becomes a critical cost for many applications as the data grows, we next examine the space overhead imposed by the various indexing schemes. The results are shown in Figure~\ref{fig:space}, where we report the space required for  10GB of raw data.

For the materialized indexes, we observe that Coconut-Tree-Full and DSTree have a smaller space overhead. Median-based solutions, such as Coconut-Tree-Full generate indexes with the leaf nodes as full as possible, whereas in prefix-based solutions there is a lot of empty space in the leaf nodes: leaves are on average 10\% full in prefix-based solutions, whereas for the median-based ones utilization reaches 97\%.
Note that in the case of Coconut-Trie-Full more space is wasted, since more leaf nodes are produced, and we cannot further compact the leaf nodes due to the specific prefix-based scheme that is used (there are 55K leaf nodes for the Coconut-Trie-Full, and 54K leaf nodes for the ADSFull). 
For the Coconut-Tree-Full, we can effectively control the number of leaf nodes produced, resulting in 6K leaf nodes with a 75\% fill rate.

For the non-materialized indexes, we can again observe the superiority of our median based solution, requiring almost half the space required by other solutions.

\Paragraph{Scalability with Data Growth}
Have identified the Coconut-Tree methods as the quickest to build data series indexes and the ADS methods as the closest contenders, we now proceed to evaluate how construction speed scales for these methods as the data size increases. We will return to the other methods when we evaluate query performance. 
In this set of experiments, we fix the amount of main memory to that of a common desktop workstation (8GB), and gradually increase the number of data series to be indexed. The results are shown in Figures~\ref{fig:fixmemory} and~\ref{fig:fixmemory+}.
We observe that when the amount of data is relatively small with respect to the available main memory, Coconut-Tree-Full and Coconut-Tree require similar times to ADSFull and ADS+, respectively.
However, as the data size increases, the random I/Os of ADSFull and ADS+ incur a significant overhead on the overall time to construct the index, and the Coconut-Tree algorithms become faster. This effect is especially pronounced for the materialized indexes in Figure~\ref{fig:fixmemory}. 
In addition, the experiments show that in Coconut-Tree-Full most of the time is spent on sorting the raw data, whereas in the case of CTree only the summarizations are sorted, and as such the external sort overhead is really small when compared to the cost of I/Os and CPU.
\begin{figure*}[tb]
	\begin{subfigure}{0.32\textwidth}
		\begin{center}
			\includegraphics[width=\textwidth]{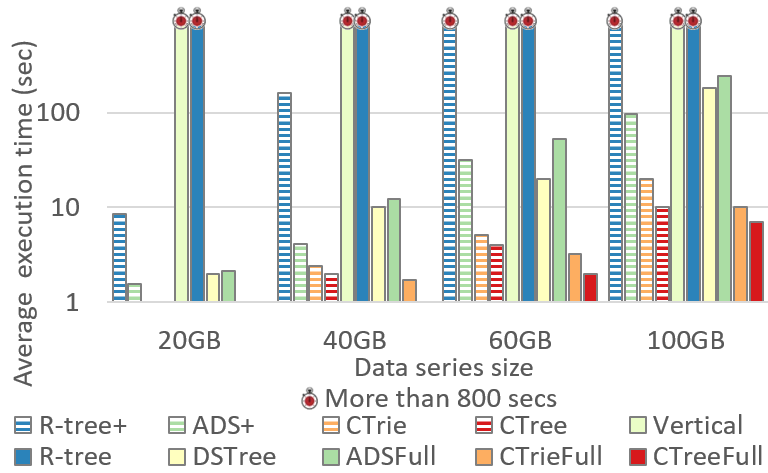}
			\caption{\blue{Exact query answering.}}
			\label{fig:exact}
		\end{center}
	\end{subfigure}
	\begin{subfigure}{0.32\textwidth}
		\begin{center}
			\includegraphics[width=\textwidth]{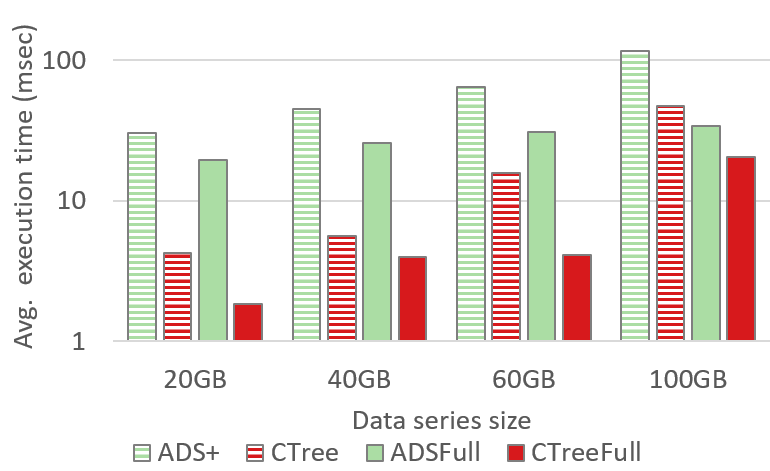}
			\caption{Approximate query answering.}
			\label{fig:approximate}
		\end{center}
	\end{subfigure}
	\begin{subfigure}{0.32\textwidth}
		\includegraphics[width=\textwidth]{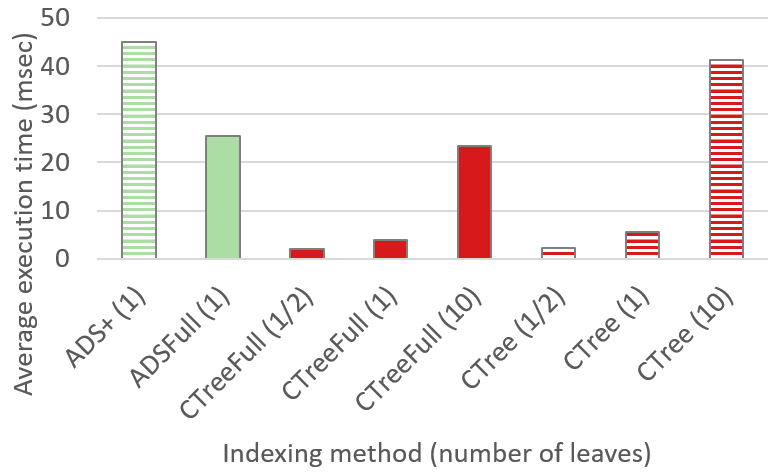}
		\caption{Approximate query answering (40G).}
		\label{fig:quality_approximate}
	\end{subfigure}
	\begin{subfigure}{0.32\textwidth}
		\includegraphics[width=\textwidth]{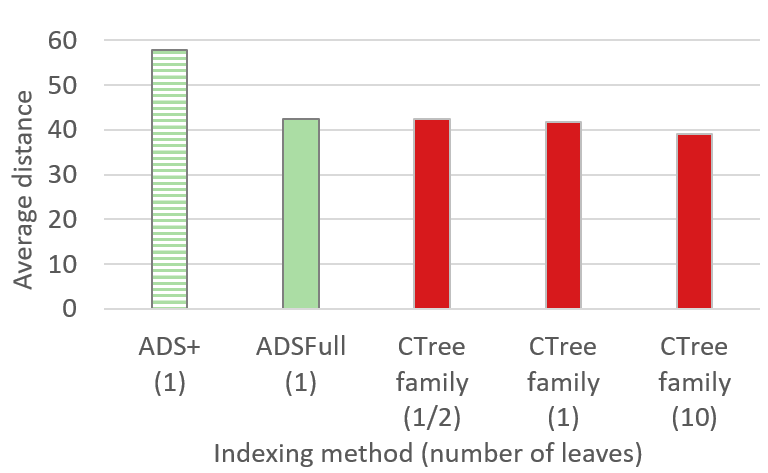}
		\caption{Average distance of approximate search.}
		\label{fig:quality_avdist}
	\end{subfigure}
	\begin{subfigure}{0.32\textwidth}
		\includegraphics[width=\textwidth]{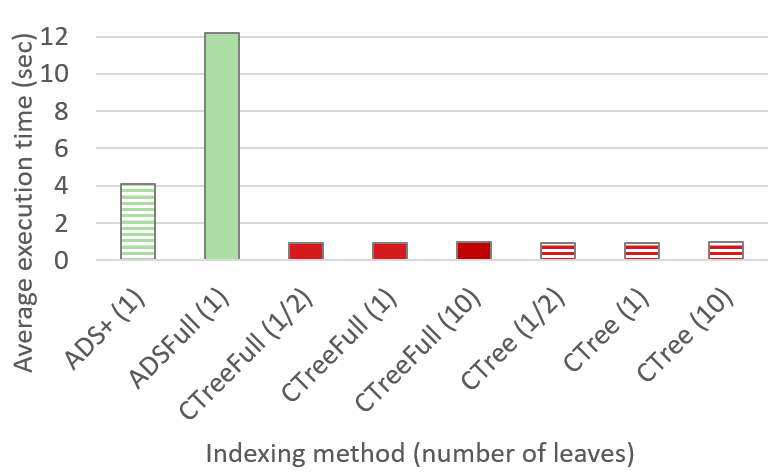}
		\caption{Exact query answering.}
		\label{fig:quality_exact}
	\end{subfigure}
	\begin{subfigure}{0.32\textwidth}
		\includegraphics[width=\textwidth]{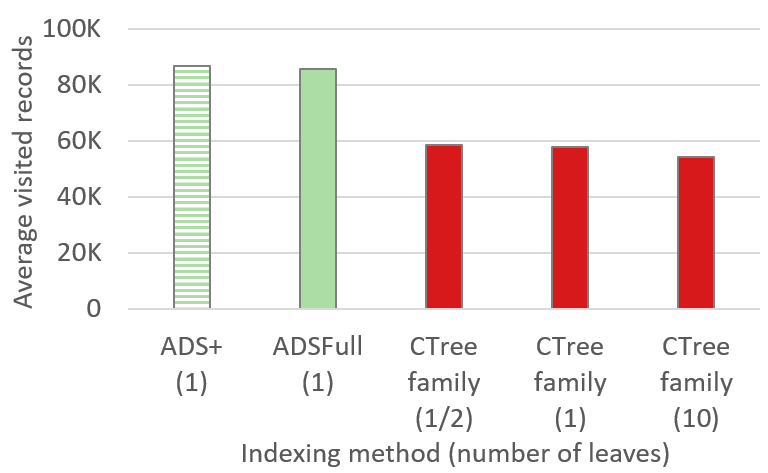}
		\caption{Visited records in exact query answering.}
		\label{fig:quality_visited records}
	\end{subfigure}
	\caption{Querying.}
\end{figure*}

\Paragraph{Variable Data series and Summary Length}

Next, we evaluate construction speed for the Coconut-Tree and ADS methods, as we vary the lengths of the individual data series that need to be indexed, and the number of segments that are used in the summarizations. We use a data series collections of 100GB, using limited memory (100K data series) for both experiments. The results for variable lengths of dataseries are shown in Figure~\ref{fig:variable}, and the results for the variable number of segments are shown in Figure~\ref{fig:variablesegments}.
When looking at Figure~\ref{fig:variable}, we observe that in all cases the Coconut-Tree variations surpass the ADS ones, demonstrating once again the superiority of Coconut-Tree in terms of construction speed. 
Regarding the variable number of segments, as we observe in Figure~\ref{fig:variablesegments}, the indexing process becomes slower when we increase the number of segments, as more segments need to be written to disk. 
We note that the ADS family does not scale beyond 16 segments as the corresponding indexing algorithms need to construct $2^{\#segments}$ root nodes in each case, and as such they have a limitation that the Coconut family has not.

\subsection{Querying}
\Paragraph{Exact Query Performance} Next we evaluate the various schemes in terms of exact query performance.
To do this, we measure execution time across 100 random exact queries as we vary the index sizes. As shown in Figure~\ref{fig:exact}, CTree and CTreeFull are faster across the board.
The 
reason is that Coconut-Tree indexes are contiguous and compact, and so fewer I/Os are needed to traverse them.

Interestingly, the non-materialized  R-tree in 40GB is faster than the materialized R-tree. This happens since R-tree+ needs only the summarizations in memory to perform query answering, whereas the materialized version needs large parts of data series, which leads to memory swapping to disk.

\Paragraph{Approximate Query Performance}
We now evaluate the performance of the different indexes in terms of approximate query answering. To do so, we measure execution time across 100 random approximate queries as we vary the size of the dataset. We focus on the indexes that were deemed most promising by the last experiment. 
The results are shown in Figure~\ref{fig:approximate}.
We observe that CTree and CTreeFull are always faster than the other methods as there are fewer nodes to traverse before reaching the target leaf node. 
In addition, the materialized versions of the indexes are faster than their non-materialized counterparts, since the records are materialized in the leaf nodes and can be directly accessed instead of issuing additional accesses to the raw data file.


\Paragraph{Approximate Query Quality vs. Performance}
In the next series of experiments, we explore whether it is possible to strike different trade-offs between performance and accuracy for approximate queries. The idea is that by searching slightly more nodes  during an approximate query and thereby sacrificing some performance, we may be able to improve accuracy by finding a better candidate. 
To run this experiment, we consider three variants of our approximate query algorithm that differ in terms of the number of nodes that get searched: half a node, a whole node, or ten adjacent nodes. Figure~\ref{fig:quality_approximate} demonstrates that approximate query execution time increases in proportion to the number of nodes we search. In Figure~\ref{fig:quality_avdist}, we measure the corresponding accuracy in terms of Euclidean distance between the search target to the closest data series we found in the searched nodes. We indeed observe in these experiments that CTree(1) (which checks one node) is more  accurate than the ADS family for 69\% of the queries, while CTree(10) is more accurate for 94\% of the queries. However, we observe that we quickly hit the point of diminishing marginal returns in terms of accuracy as we search more nodes. 

Since the first step of the exact search is the execution of an approximate query, we might expect that a better initial approximate result would lead to more pruning and thus improved performance for exact queries.
Figure~\ref{fig:quality_visited records} indeed shows that the ADS family on average  visits more than 80K records during exact query answering, whereas the Coconut family visits fewer than 59K records in all cases.
In Figure \ref{fig:quality_exact}, however, we observe that all the Coconut variants perform approximately the same. This implies that the performance improvement that we observe for the Coconut family compared with the ADS family mostly arises due to the compactness and contiguity of the Coconut indexes, which allow us to issue fewer I/Os during exact queries.

\subsection{Complete Workloads on Real Datasets}

\begin{figure*}[tb]
	\begin{subfigure}{0.5\textwidth}
		\includegraphics[width=\textwidth]{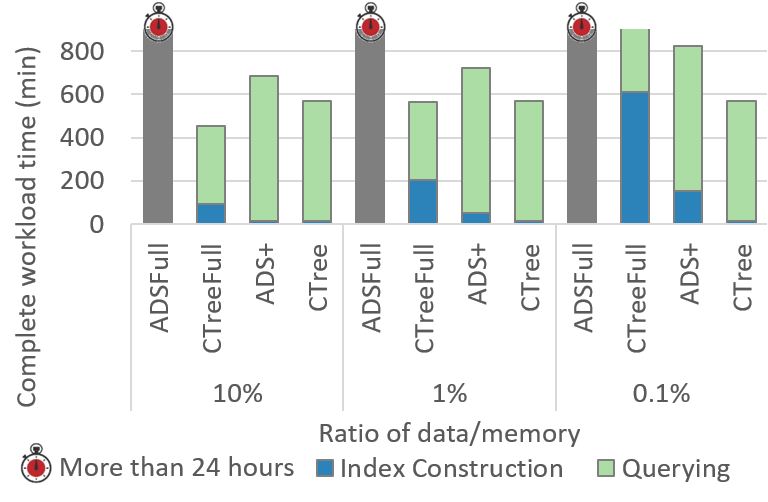}
		\caption{Astronomy - complete workload.}
		\label{fig:real_astro}
	\end{subfigure}
	\begin{subfigure}{0.5\textwidth}
		\includegraphics[width=\textwidth]{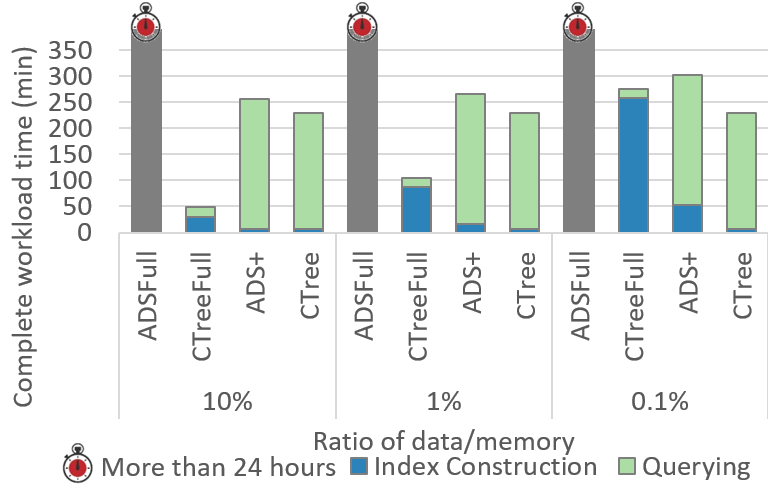}
		\caption{Seismic - complete workload.}
		\label{fig:real_seismic}
	\end{subfigure}
	\caption{Complete Workloads.}
\end{figure*}

We now compare Coconut to the state-of-the-art, simulating the complete process of index construction and query answering.
The results are shown in Figure~\ref{fig:real_astro} for the Astronomy dataset and in Figure~\ref{fig:real_seismic} for the Seismic dataset.

The index sizes for the astronomy dataset were as follows: ADSFull: 311GB, ADS+: 19GB, CTree: 10GB, CTreeFull: 298GB; and for the seismic dataset: ADSFull: 111GB, ADS+: 6GB, CTree: 4GB, CTreeFull: 108GB.

We measure the time to construct first the corresponding indexes, and then to answer 100 exact queries over the constructed index, using various memory configurations.
As shown, when we constrain the available memory, Coconut-Tree becomes better in all cases, for both the materialized and non-materialized approaches, corroborating the experimental results with the synthetic datasets.
An interesting observation here is that the queries are harder on these datasets for all indexes, because the datasets were denser (for a detailed discussion on hardness see~\cite{ZoumpatianosLPG15}). 
As a result, pruning was not as efficient as with the random walk data. 
Therefore, even though Coconut was faster than all competing methods, it still had to scan a considerable amount of data in order to answer the exact queries.


\subsection{Insertions} \label{sec:exp_insertions}

\begin{figure}[tb]
	\includegraphics[width=\columnwidth]{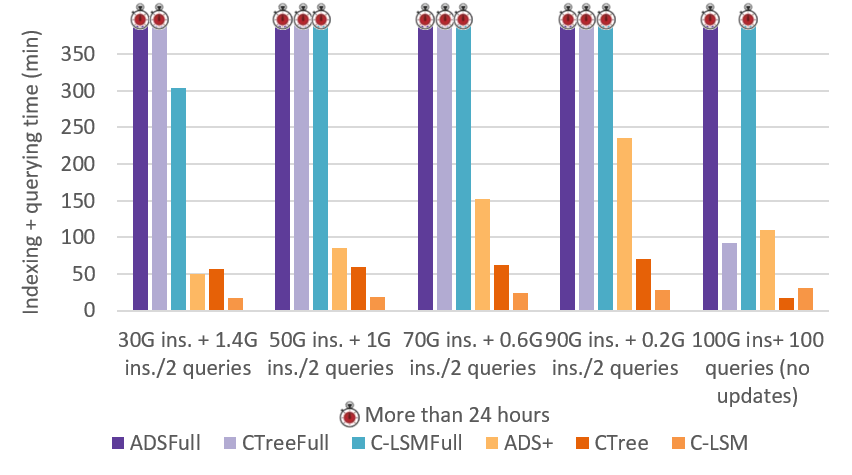}
	\caption{Updates for both materialized and non-materialized versions of ADS, Coconut-Tree and Coconut-LSM.}
	\label{fig:updatess}
\end{figure}

Next, we evaluate the different indexes  in the presence of insertions of new data series. We focus on the ADS and Coconut-Tree families as they were shown to perform best for index construction. This time, we also include Coconut-LSM in the experiment (i.e., as the structural difference between Coconut-LSM and Coconut-Tree only manifest themselves in the presence of insertions). In particular, we use C-LSM and C-LSM-Full as non-materialized and materialized instances of Coconut-LSM, respectively. We use a synthetic workload consisting of 100 random exact queries, where every two queries are interleaved by a batch of insertions. We control the experiment by ensuring that the final data size after all insertions at the end of each of the experiments is 100GB, while the initial data size and the insertion batch size vary. In addition, we limit the available memory to 0.01\% of the data size. The results in Figure~\ref{fig:updatess} show that in the presence of insertions, C-LSM performs at least twice as fast as the other approaches. The reasons are that (1) the LSM-tree on top of which C-LSM is built optimizes heavily for insertion workloads by buffering and later sort-merging data and thereby using only sequential rather than random writes, and (2) C-LSM is non-materialized and so only new incoming summarizations get indexed while the bulk of the data (i.e., the data series) are appended to the raw file. We further observe that in the absence of insertions (the final set of bars), C-LSM and CTree perform similarly because in this case both consist of one contigous and compact level of summarizations. We attribute the performance difference in this case to implementation differences between BerkeleyDB and RocksDB, on top of which they are implemented. C-LSM-Full does not perform as well as CTree due to the overheads of continually sort-merging the whole data rather than just the summarizations. Overall, we observe here again that being able to sort the data allows us to optimize for different workload characteristics (in this case for insertions), as well as to introduce Coconut-LSM as the first highly write-optimized data series index. 

\subsection{Sliding windows}

\begin{figure*}[tb]
	\centering\includegraphics[width=0.8\textwidth]{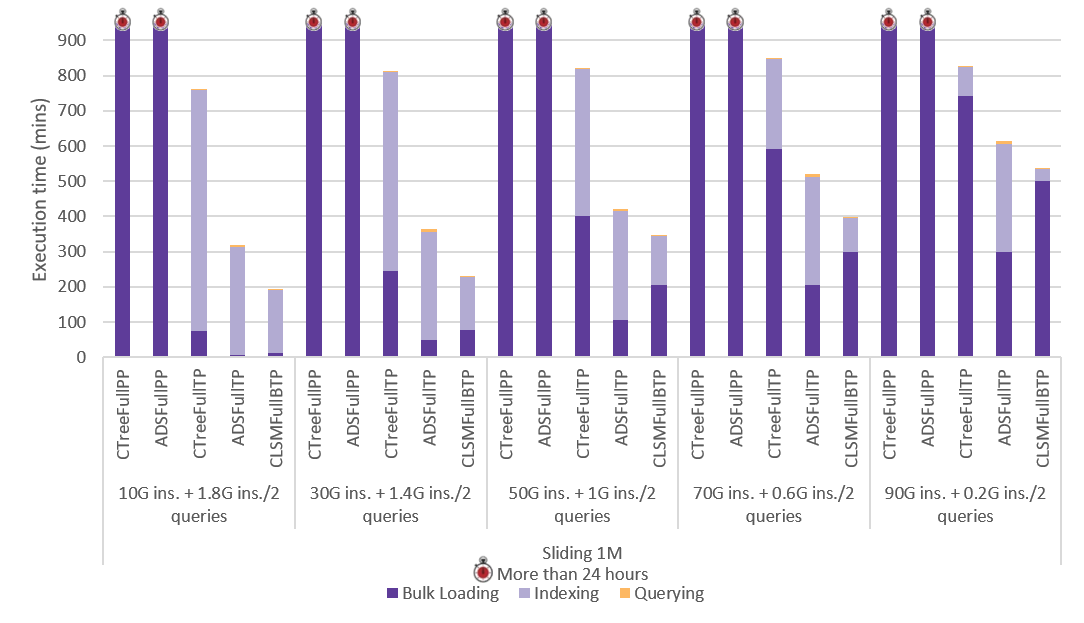}
	\caption{Sliding window experiments with fixed length window (materialized methods).}
	\label{fig:ssliding}
\end{figure*}

\begin{figure}[tb]
	\includegraphics[width=\columnwidth]{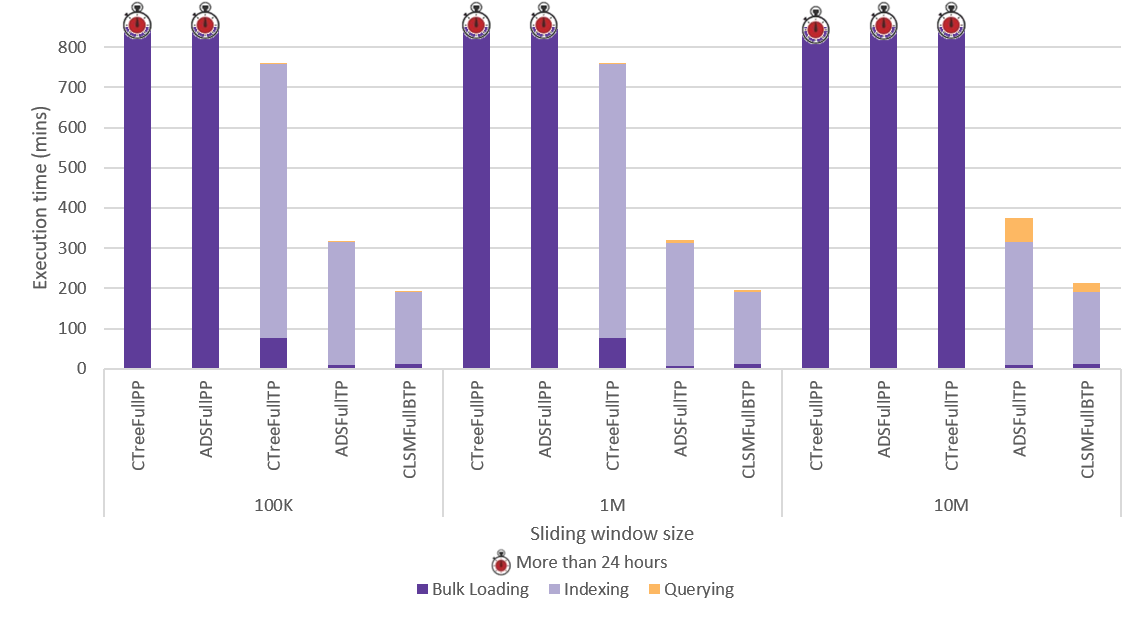}
	\caption{Sliding window experiments with variable length window (materialized methods). }
	\label{fig:varssliding}
\end{figure}

\begin{figure*}[tb]
	\centering\includegraphics[width=0.8\textwidth]{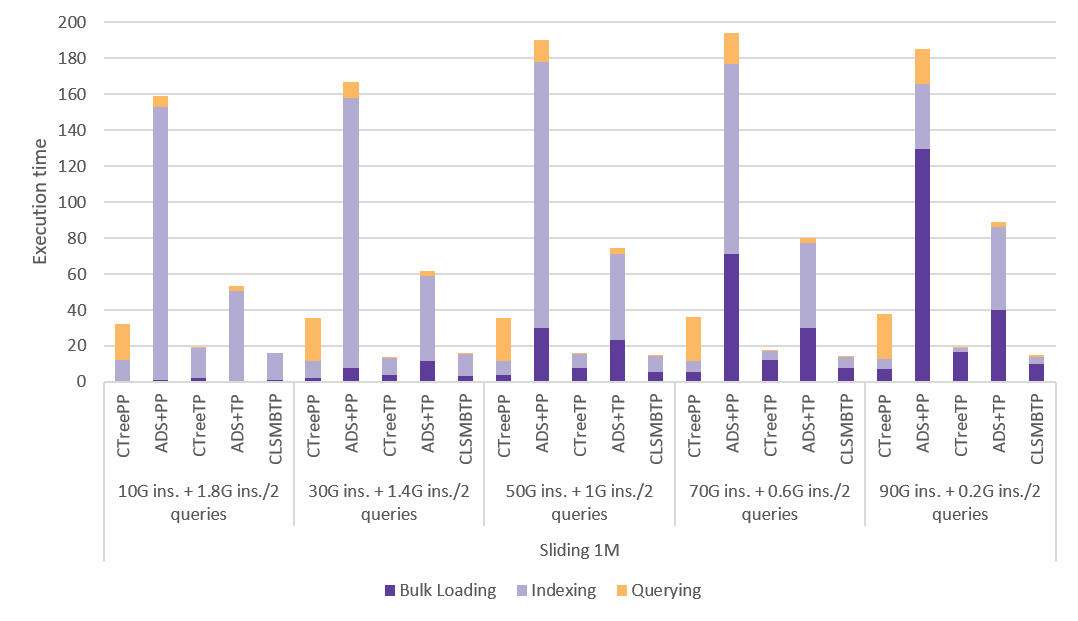}
	\caption{Sliding window experiments with fixed length window (non-materialized methods).}
	\label{fig:sslidingplus}
\end{figure*}

\begin{figure}[tb]
	\includegraphics[width=\columnwidth]{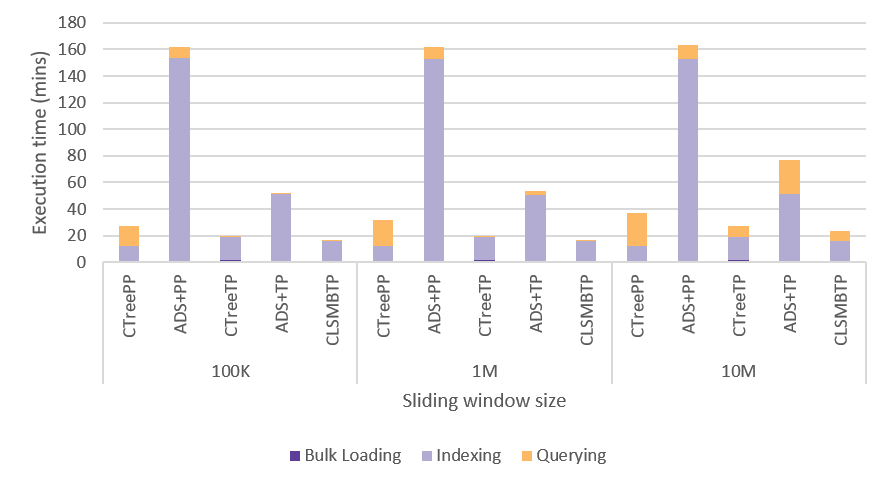}
	\caption{Sliding window experiments with variable length window (non-materialized methods). }
	\label{fig:varsslidingplus}
\end{figure}

Finally, we show that sortable  data series summarizations further allow to process efficient sliding window queries for streaming applications. To recap from Section \ref{sec:sliding}, the baseline approaches for processing sliding window queries with unsortable summarizations are post-processing (PP) and temporal partitioning (TP). Post-Processing (PP) performs a regular exact query over the whole index and discards data series based on their creation timestamp after they are retrieved from storage. Temporal Partitioning (TP), on the other hand, creates a separate temporal partition for every new batch of insertions thereby allowing queries to ignore partitions with older data than the specified window. Our proposed approach, Bounded Temporal Partitioning (BTP), creates temporal partitions as with TP, but it also sort-merges partitions as they grow older. This allows to restrict the overall number of partitions. 
We implemented the first two approaches for Coconut-Tree-Full and ADSFull, and we call them CTreeFullPP, CTreeFullTP, ADSFullPP and ADSFullTP. We implemented the BTP approach on top of Coconut-LSM-Full, and we refer to this algorithm as CLSMFullBTP. 
We conduct the experiment as in Section \ref{sec:exp_insertions} by interleaving batches of insertions with exact queries, but now each of the queries is an exact sliding window query over the most recent one million data series. The final data size after all insertions is 100GB, and the memory we use is $0.1\%$ of the final data size. 

Figures~\ref{fig:ssliding} and~\ref{fig:sslidingplus} show the experimental results for the materialized and non-materialized indexes, respectively. The PP approach is slowest because it accesses the most data. We stopped the execution of all the PP methods after 24 hours. The TP approach performs better than PP because it allows to restrict the search to the most recent temporal partitions. However, the high number of partitions leads to random I/O across partitions. Furthermore, TP does not enable effective pruning within each of the partitions because the search starts from scratch for each partition, and so it cannot leverage the lower-bounding property of invSAX as effectively to spatially prune within each of the partition. BTP, on the other hand, performs best in all cases because it further sort-merges partitions to restrict their number and to create large, compact and contiguous partitions for older data. Thus, this approach allows us to prune more at older partitions, and it makes the access patterns to disk less random and more skip-sequential. 

Figures~\ref{fig:varssliding} and ~\ref{fig:varsslidingplus} repeat the experiments as we vary the sliding window size for the materialized and non-materialized indexes, respectively. For this experiment, we start with 10GB of data and each insertion batch is 1.4GB. We observe that querying takes longer with larger window sizes as a larger fraction of the data has to be accessed. In all cases, however, BTP continues to dominate the other approaches. Overall, this demonstrates that sortable summarizations provide us with more scalable means of analyzing data at different temporal granularities, an important property for modern data-heavy streaming applications.

\section{Conclusions and Future Work}

In this paper, we show that state-of-the-art data series indexes do not scale well for massive data sizes in terms of performance for index construction, updating and querying. We show that the reason is that existing data series summarizations, on top of which these indexes are built, are unsortable. As a result, such indexes are constructed and updated through expensive top-down insertions that create a non-contiguous index that is expensive to query. 
To alleviate this problem, we propose the first \emph{sortable} data series summarizations, showing that indexing based on sortable summarizations optimizes both indexing and querying.
We start by creating and exploring a prefix-based bottom-up indexing algorithm,
which merely solve the problem of data contiguity.
We proceed 
by exploring median-based split trees, and showing that this approach outperforms the state-of-the-art for both index construction and querying time.
Among the benefits of the approach is that the resulting index structure is balanced, providing guarantees on query execution time.
Moreover, we design the first write-optimized data series index by using log-structured updates, a technique that is enabled by having sortable data series summarizations. 
Finally, we explore three approaches for query answering over streaming sets and we provide an efficient solution in this direction.
As future work, we intend to explore how Cococut can be parallelized, by exploring parallel UB-Tree index building algorithms.



\section{Acknowledgements}
This project has received funding from the European Union’s Horizon 2020 research and innovation programme under the Marie Skłodowska-Curie grant agreement No 748945.


\bibliographystyle{spmpsci}      
\bibliography{Coconut_VLDBJ} 
%
%

\end{document}